\documentclass[12pt]{article}
\usepackage{bm}

\usepackage[titletoc,title]{appendix}

\usepackage[T1]{fontenc}

\newfont{\elevenpt}{ptmr8t at 11pt}

\usepackage{hyperref}

\DeclareBoldMathCommand{\cmd}{<math>}

\usepackage{amssymb}
\usepackage{amsfonts}
\usepackage{amsmath}
\usepackage{graphicx}
\usepackage{float}
\def\therefore{\hbox{.\raise1ex\hbox{.}.}}
\def\because{\leavevmode\raise1ex\hbox{.}.\raise1ex\hbox{.}}

\def\qed{\quad \rule{0.5em}{2ex}}
\def\co{\mathop{\mbox{\rm co}}}             
\newcommand{\R}{\mathbb{R}}

\newcommand{\bw}{\boldsymbol{w}}

\newcommand{\bu}{\boldsymbol{u}}

\newcommand{\bone}{\boldsymbol{1}}

\newcommand{\vbx}{\mathbf{x}}
\newcommand{\vby}{\mathbf{y}}
\newcommand{\vbz}{\mathbf{z}}

\def\qed{\quad\rule{0.5em}{2ex}}

\newtheorem{Theorem 1}{Theorem 1}[section]

\usepackage{authblk}

\begin{document}

\title{Cautious and fair social evaluation under risk\footnote{The authors are grateful to Chiaki Hara, Youichiro Higashi, Leo Kurata, Satoshi Nakada, Jawwad Noor, Marcus Pivato, Takuma Wakayama, Shohei Yanagita, Michael Zierhut, and the seminar participants at the Online Social Choice and Welfare Seminars and the Kyoto Institute of Economic Research. Grants from JSPS KAKENHI (No. 20K01564, 25K05006) are gratefully acknowledged.}}
\author{Kaname Miyagishima\footnote{Hitotsubashi University. Email: Kaname1128@gmail.com}
  \hspace{1mm}and
Kensei Nakamura\footnote{Hitotsubashi University. Email: kensei.nakamura.econ@gmail.com}}
\date{September 11, 2026}
\maketitle

\begin{abstract}
  In this paper, we study criteria for social evaluation of resource distributions under risk when individuals have expected utility preferences but their utility functions are interpersonally non-comparable.
  In particular, we focus on cautious evaluation with respect to ex-post distributions.
  As a key rationality axiom, we adopt Negative Certainty Independence, introduced by Dillenberger (2010). Using this axiom together with Pareto conditions that are compatible with both the rationality axiom and ex-post equity, we characterize a class of social criteria called Cautious Equally Distributed Equivalent (CEDE) criteria.
  These criteria evaluate risky situations by the lowest certainty equivalent over a set of social utility functions constructed from individual utility functions.
\end{abstract}

\section{Introduction}
Social evaluations under risk often require a balance between caution about adverse ex-post outcomes and willingness to undertake socially beneficial risks.
For instance, the adoption of a new technology may generate substantial aggregate benefits while exposing some individuals to unemployment, poverty, or other serious losses and thereby exacerbating ex-post inequality (e.g., Acemoglu and Johnson 2023).
A social evaluator should therefore be cautious about unfavorable distributions of final outcomes.
At the same time, excessive caution may lead society to reject risks that are likely to produce widely shared benefits with little chance of causing serious losses.
The question is how social evaluation can accommodate both concerns.

We study this question in a framework where outcomes are allocations of monetary payoffs and individuals have possibly heterogeneous expected-utility preferences over risky monetary prospects.
Individual preferences are interpersonally non-comparable.
We consider the aggregation of individuals' risk preferences while taking the ex-post distribution of payoffs into account.
We derive social criteria that address the above question from requirements of rationality, ex-post equity, and efficiency.

Our central axiom is Negative Certainty Independence (NCI), introduced by Dillenberger (2010).
NCI requires that if a risky social prospect $p$ is weakly preferred to a riskless allocation $\vbx$, then mixing both alternatives with the same third prospect $r$ cannot reverse this ranking.
The condition is clearly weaker than von Neumann--Morgenstern Independence.
Dillenberger (2010) uses NCI to capture the certainty effect (Kahneman and Tversky 1979), and Cerreia-Vioglio et al. (2015) show that it yields what they call a {\it cautious expected utility} representation.
In the present setting, NCI permits caution toward serious outcomes without requiring the social evaluator to reject every beneficial risk.
NCI can also be interpreted in terms of dynamic consistency.
If $p$ is a sufficiently attractive risk relative to the riskless allocation $\vbx$, NCI requires choosing the mixture of $p$ and $r$ over the corresponding mixture of $\vbx$ and $r$; if the possibility of $r$ is subsequently ruled out, society proceeds with the preferred prospect $p$.
Thus, NCI is well suited to our purpose of combining caution with a willingness to choose beneficial risks.
Appendix A shows that, under Consequentialism and Reduction of Two-stage Lotteries, NCI is equivalent to a certainty-restricted form of dynamic consistency, following an argument similar to that of Karni and Schmeidler (1991).
Furthermore, as discussed later, NCI is compatible with a certain Pareto condition that is violated under von Neumann--Morgenstern (vNM) Independence,
and hence the weaker rationality condition allows us to respect individuals' preferences more strongly.

Our first result clarifies the consequences of combining NCI with Ex-ante Pareto.
We show that Continuity, Ex-ante Pareto, and NCI characterize, at every preference profile, a linear aggregation of individual expected utilities.
Thus, even though NCI is significantly weaker than vNM Independence, a linear-aggregation result analogous to that of Harsanyi (1955) survives, and genuinely cautious aggregation through multiple social risk preferences is ruled out.
This result also exposes a conflict with ex-post equity: even a weak transfer requirement is incompatible with Ex-ante Pareto under NCI.
Moreover, on the full domain of risk preferences, adding Invariance on Degenerate Lotteries (IDL), which requires social comparisons of riskless allocations to be independent of individuals' risk attitudes, leads to dictatorship.
These results indicate that Ex-ante Pareto is too demanding for the cautious and ex-post equitable approach pursued here.

Accordingly, we replace Ex-ante Pareto with two restricted requirements: Ex-post Pareto, which applies to riskless allocations, and Pareto for Equal Risk, which applies when every realized allocation is equal (Fleurbaey 2010).
Our second result introduces a novel class of social criteria called {\it Cautious Equally Distributed Equivalent} (CEDE) criteria and characterizes this class using these weaker Pareto conditions and NCI.
A CEDE criterion is constructed in the following three steps.
First, each ex-post allocation is replaced by its equally distributed equivalent (EDE) under the social ordering.
This is the common payoff that, if received by every individual, makes the resulting equal allocation socially indifferent to the original allocation (Atkinson 1970).
Second, each prospect is transformed into a socially indifferent prospect, called its EDE prospect, by replacing each allocation in its support with its EDE allocation, while preserving the original probabilities.
Third, this prospect is evaluated by its lowest certainty equivalent over a set of social utility functions constructed as convex combinations of individual utility functions.
As in Cerreia-Vioglio et al. (2015), caution consists in evaluating the prospect according to the lowest certainty equivalent across the set of social utility functions.
As in the EDE-based approach of Fleurbaey and Zuber (2017), CEDE criteria evaluate social risk through a prospect formed from the EDE values of ex-post allocations.
An important observation is that social vNM functions are constructed as convex combinations of individual ones, as in Harsanyi's (1955) aggregation theorem.
This observation is our key technical contribution: we extend Harsanyi's linear-aggregation result to nonlinear social preferences.
To the best of our knowledge, such an extension has not previously been established in the literature.
This observation is also useful for obtaining the impossibility results.

Our third result concerns a stronger Pareto requirement called Pareto for Equal or No Risk and a more specific CEDE criterion.
As shown by Fleurbaey and Zuber (2017), this requirement is, in general, incompatible with ex-post equity under a social expected-utility representation on an unrestricted domain of individual risk preferences.
In contrast, some CEDE criteria are compatible with Pareto for Equal or No Risk even on the unrestricted domain.
We specify a subclass of CEDE criteria satisfying Pareto for Equal or No Risk.
Furthermore, we show that Continuity, Pareto for Equal or No Risk, NCI, IDL, and a weak ex-post transfer axiom characterize the maximin CEDE criterion.
This criterion first evaluates each allocation by the payoff of the worst-off individual and then evaluates the resulting equal prospect by the lowest individual certainty equivalent.

The paper is organized as follows.
Section 2 presents the model.
Section 3 introduces NCI and discusses its interpretation.
Section 4 studies Ex-ante Pareto under NCI and establishes the linear-aggregation and dictatorship results.
Section 5 characterizes CEDE criteria using the two restricted Pareto conditions.
Section 6 studies Pareto for Equal or No Risk and characterizes the maximin CEDE criterion.
Section 7 provides concluding comments.
The appendices relate NCI to dynamic consistency, contain the proofs, and establish the independence of the axioms.

\section{The Model}
We consider the problem of evaluating allocations of monetary payoffs under risk.
Let $N=\{1,\ldots,n\}$ be the set of individuals.
We write $\Delta(N)=\left\{\boldsymbol{\alpha}\in\R_+^N\ \middle|\ \sum_{i\in N}\alpha_i=1\right\}$.
The set of monetary payoffs is the bounded interval $X\equiv [\underline{x},\overline{x}]\subset \R$, where $\underline{x}<\overline{x}$.
The set of allocations is $\mathcal X\equiv X^N$.
Generic elements of $\mathcal X$ are denoted by $\vbx$, $\vby$, and $\vbz$,
and individual $i$'s monetary payoffs are denoted by $x_i$, $y_i$, and $z_i$.

Let $\Delta(\mathcal X)$ denote the set of all Borel probability measures on $\mathcal X$, endowed with the topology of weak convergence.
We refer to the elements of $\Delta(\mathcal X)$ as lotteries or prospects and denote generic elements by $p$, $q$, and $r$.
Let us denote the degenerate lottery that yields the allocation $\vbx \in \mathcal X$ with probability $1$ by $\delta(\vbx)$.
The set of degenerate lotteries is denoted by $\Delta^d(\mathcal X)$.
Let $\Delta^e(\mathcal X)$ denote the set of lotteries under which all individuals receive equal monetary payoffs ex post.
Such lotteries are called equal lotteries.
Formally, each equal lottery has support contained in $\{x\bone\in \mathcal X\mid x\in X\}$.
Generic elements of $\Delta^e(\mathcal X)$ are denoted by $p^e$, $q^e$, and $r^e$.

We assume that each individual is an expected utility maximizer and cares only about her or his own monetary payoff.
For each $p \in \Delta(\mathcal X)$, the expected utility of individual $i$ is
\begin{align*}
  Eu_i(p)=\int_{\mathcal X}u_i(x_i)\,dp(\vbx),
\end{align*}
where $u_i:X \rightarrow \R$ is a continuous and strictly increasing vNM utility function.\footnote{All our results remain valid if the domain is further restricted to concave vNM utility functions.}
We normalize it so that $u_i(\underline{x})=0$ and $u_i(\overline{x})=1$.
Let $\mathcal{U}$ denote the set of such vNM functions.
Since a vNM utility representation is unique up to a positive affine transformation, these two endpoint conditions select a unique normalized representation without loss of generality.
The certainty equivalent of $p$ with respect to $u_i \in \mathcal{U}$ is denoted by $c(p, u_i)=u_i^{-1}(Eu_i(p))$.

Given a domain $\mathcal{D}\subseteq\mathcal{U}^N$,
a {\it social ordering function} $\succsim$ maps a profile of individual vNM functions $\bu=(u_1,\ldots,u_n) \in \mathcal{D}$ into a complete and transitive binary relation $\succsim^{\bu}$ on $\Delta(\mathcal X)$.
For all $\bu \in \mathcal{D}$ and all $p, q \in \Delta(\mathcal X)$, $p \succsim^{\bu} q$ is interpreted as ``$p$ is socially weakly preferred to $q$''.
The asymmetric and symmetric parts of $\succsim^{\bu}$ are denoted by $\succ^{\bu}$ and $\sim^{\bu}$, respectively.
We impose the following standard continuity condition.
\begin{description}
  \item[Continuity.] For all $\bu \in \mathcal{D}$ and all $q \in \Delta(\mathcal X)$,
    $\{p\in \Delta(\mathcal X)\mid p \succsim^{\bu} q \}$ and $\{p\in \Delta(\mathcal X)\mid q \succsim^{\bu} p \}$ are closed.
\end{description}

\section{Negative Certainty Independence}
In this section, we introduce our central rationality requirement, Negative Certainty Independence, and discuss its meaning and justification.
The axiom was introduced by Dillenberger (2010).
\begin{description}
  \item[Negative Certainty Independence (NCI).] For all $\bu \in \mathcal{D}$, all $\delta(\vbx), p, r \in \Delta(\mathcal X)$, and all $\lambda \in (0, 1)$, $p \succsim^{\bu} \delta(\vbx)$ implies
    $\lambda p + (1-\lambda)r \succsim^{\bu} \lambda \delta(\vbx)+ (1-\lambda)r $.
\end{description}
This axiom states that if a riskless prospect $\delta(\vbx)$ is weakly worse for society than a prospect $p$, then making $\delta(\vbx)$ risky by mixing it with another prospect $r$ eliminates its certainty appeal, and thus the mixture of $\delta(\vbx)$ should also be weakly worse than the mixture of $p$ with the same prospect $r$.
Cerreia-Vioglio et al. (2015) argue that NCI captures the certainty effect that explains Allais-type behavior (Kahneman and Tversky 1979).
This refers to the commonly observed pattern of behavior in which individuals put more weight on certainty than expected utility theory predicts.
Allais argues that this tendency is a form of prudent behavior (Allais 1953; Mongin 2019).

NCI can also be interpreted as a dynamic consistency condition weaker than the von Neumann--Morgenstern independence axiom stated below.
\begin{description}
  \item[Independence.] For all $\bu \in \mathcal{D}$, all $p, q, r \in \Delta(\mathcal X)$, and all $\lambda \in (0, 1)$, $p \succsim^{\bu} q$ if and only if
    $\lambda p + (1-\lambda)r \succsim^{\bu} \lambda q+ (1-\lambda)r $.
\end{description}
When mixed lotteries are considered as two-stage lotteries, if $p$ is at least as good for society as $q$ in the second stage, then it would be reasonable to think that $\lambda p + (1-\lambda)r$ is socially at least as good as $\lambda q + (1-\lambda)r$ in the first stage as well.
It is well recognized that Independence is an appealing normative condition of rationality in the sense of dynamic consistency (Karni and Schmeidler 1991).
As argued in Appendix A, NCI is a relaxation of this rationality requirement.
If a lottery $p$ is socially at least as good as a riskless lottery $\delta(\vbx)$ in the second stage, it is natural to require that $\lambda p + (1-\lambda)r$ be at least as good as $\lambda \delta(\vbx) + (1-\lambda)r$ in the first stage as well.
This relaxation of Independence would be acceptable for a cautious evaluator.
If a risky prospect $p$ is good enough to compensate for giving up a riskless prospect $\delta(\vbx)$, the evaluator would confidently assess that $\lambda p + (1-\lambda)r$ is at least as good for society as $\lambda \delta(\vbx) + (1-\lambda)r$ in the first stage, because the certainty appeal is eliminated in the latter prospect.
Thus, NCI is a reasonable dynamic consistency requirement for cautious evaluation that does not rule out beneficial risks.

One might object to NCI for reasons similar to Diamond's (1967) critique of social expected utility in Harsanyi (1955).
For instance, let $N=\{1, 2\}$, $\vbx=(1,0)$, and $\vby=(0,1)$.
Ex-ante fairness may require $p=\frac12\delta(\vbx)+\frac12\delta(\vby)\succ\delta(\vbx)$ because $p$ gives the individuals equal chances of receiving the payoff of 1.
Let $r=\frac14\delta(\vbx)+\frac34\delta(\vby)$ and $\lambda=\frac13$.
Then $\lambda p+(1-\lambda)r=\frac13\delta(\vbx)+\frac23\delta(\vby)$, whereas $\lambda\delta(\vbx)+(1-\lambda)r=\frac12\delta(\vbx)+\frac12\delta(\vby)$.
Given $p\succ\delta(\vbx)$, NCI requires the former prospect to be weakly preferred to the latter.
On the other hand, ex-ante fairness may instead favor the latter prospect because it gives the individuals equal chances, contrary to NCI.
However, ex-ante fairness is only one of several relevant normative values.
The main purpose of this paper is to explore social welfare criteria that are rational and cautious and that respect ex-post equality.
As discussed above, NCI is well suited as a rationality requirement for this purpose.
Moreover, ex-ante fairness may conflict not only with NCI but also with ex-post equality.
As Adler and Sanchirico (2006) and Mongin and Pivato (2016) explain, equalizing individuals' ex-ante prospects may favor a lottery even when it generates more unequal ex-post distributions, whereas an ex-post criterion places greater weight on avoiding outcomes that leave even a small number of individuals severely disadvantaged.
To illustrate this conflict, let $N=\{1,2\}$, $X=[0,4]$, and $u_1(x)=u_2(x)=x$, and consider
\begin{align*}
  p=\frac12\delta((4,0))+\frac12\delta((0,4)) \ \text{ and } \
  q=\delta((3,1)).
\end{align*}
Under $p$, each individual has an equal chance of receiving the high payoff $4$, whereas under $q$, individual 1 receives a higher payoff than individual 2 with certainty.
Thus, ex-ante fairness may favor $p$ because it gives the individuals equal chances of occupying the advantaged position.
However, $q$ raises the lowest realized payoff from $0$ to $1$ and is more equal than every allocation in the support of $p$.
Thus, giving priority to ex-ante fairness can make it difficult to respond adequately to ex-post poverty and inequality.
A substantial body of literature similarly gives priority to ex-post considerations, including Adler and Sanchirico (2006), Hammond (1981), Fleurbaey (2010), Grant et al. (2012), and Fleurbaey and Zuber (2017).
Building on this approach, this paper studies social evaluation criteria that combine ex-post equity with NCI, reflecting our concern for both dynamic consistency and cautious evaluation under risk.

\section{Ex-ante Pareto under Negative Certainty Independence}
We now examine the implications of combining NCI with the standard Ex-ante Pareto principle.
\begin{description}
  \item[Ex-ante Pareto.] For all $\bu \in \mathcal{D}$ and all $p, q \in \Delta(\mathcal X)$, if $Eu_i(p) > Eu_i(q)$ for all $i \in N$, then $p \succ^{\bu} q$.
\end{description}
This axiom requires that if all individuals unanimously prefer $p$ to $q$, then $p$ should be socially preferred to $q$ as well.
Although this is a standard efficiency condition under risk, its combination with NCI has restrictive implications for cautious and ex-post equitable social evaluation.

The following proposition shows that a continuous social ordering function that satisfies NCI and Ex-ante Pareto is characterized by a linear aggregation of individual expected utilities at every preference profile.
That is, a linear-aggregation result analogous to Harsanyi's (1955) theorem continues to hold even when Independence is weakened to NCI.
\begin{description}
  \item[Proposition 1.] A social ordering function $\succsim$ satisfies Continuity, Ex-ante Pareto, and NCI if and only if, for every $\bu\in\mathcal D$, there exist coefficients $(a_i(\bu))_{i\in N}\in\Delta(N)$ such that, for all $p,q\in\Delta(\mathcal X)$,
    \begin{align*}
      p\succsim^{\bu}q
      \quad\Longleftrightarrow\quad
      \sum_{i\in N}a_i(\bu)Eu_i(p)
      \geq
      \sum_{i\in N}a_i(\bu)Eu_i(q).
    \end{align*}
\end{description}
Thus, Ex-ante Pareto restores an expected-utility representation and rules out genuinely cautious aggregation rules, such as those based on cautious expected utility theory, which represents caution through multiple social utility functions.

One implication of Proposition 1 is that, under NCI, Ex-ante Pareto is incompatible even with weak forms of ex-post equity.
For instance, consider two individuals with $u_1(x)=u_2(x)=(x-\underline{x})/(\overline{x}-\underline{x})$, and let $m=(\underline{x}+\overline{x})/2$.
By Proposition 1, the social ordering is represented by $a_1x_1+a_2x_2$ for some $a_1,a_2 \in \R_+$ such that $a_1+a_2=1$.
An ex-post equitable evaluation would require
\begin{align*}
  \delta(m\bone)\succ^{\bu}\delta\big((\overline{x},\underline{x})\big)
  \quad\text{and}\quad
  \delta(m\bone)\succ^{\bu}\delta\big((\underline{x},\overline{x})\big).
\end{align*}
The linear-aggregation representation would then imply
\begin{align*}
  m(a_1+a_2)>a_1\overline{x}+a_2\underline{x}
  \quad\text{and}\quad
  m(a_1+a_2)>a_1\underline{x}+a_2\overline{x}.
\end{align*}
Adding these inequalities gives
$2m(a_1+a_2)>(\underline{x}+\overline{x})(a_1+a_2)$,
which contradicts $2m=\underline{x}+\overline{x}$.
Thus, under NCI, Ex-ante Pareto is incompatible with ex-post equity.\footnote{It is not difficult to directly show the incompatibility between Ex-ante Pareto, NCI, and ex-post equity without using Proposition 1.}

We next introduce a condition requiring social evaluations of riskless allocations to be independent of individual risk preferences.
Chambers and Echenique (2012) introduce an analogous invariance axiom for constant allocations on a broader domain of risk preferences that need not satisfy expected utility. IDL is its counterpart in the present setting.
\begin{description}
  \item[Invariance on Degenerate Lotteries (IDL).] For all $\bu,\bu'\in\mathcal D$ and all $\delta(\vbx),\delta(\vby)\in\Delta^d(\mathcal X)$,
    \begin{align*}
      \delta(\vbx)\succsim^{\bu}\delta(\vby)
      \Longleftrightarrow
      \delta(\vbx)\succsim^{\bu'}\delta(\vby).
    \end{align*}
\end{description}
IDL is natural because risk attitudes are irrelevant when only riskless allocations are compared.

When IDL is combined with Ex-ante Pareto under the full-domain assumption $\mathcal D=\mathcal U^N$, the linear-aggregation representation in Proposition 1 collapses further to dictatorship.
We call a social ordering function {\it dictatorial} if there exists $d\in N$ such that, for every $\bu \in \mathcal D$ and all $p,q\in\Delta(\mathcal X)$,
\begin{align*}
  p\succsim^{\bu}q
  \Longleftrightarrow
  Eu_d(p)\geq Eu_d(q).
\end{align*}

\begin{description}
  \item[Proposition 2.] Suppose that $\mathcal D=\mathcal U^N$. A social ordering function $\succsim$ satisfies Continuity, Ex-ante Pareto, NCI, and IDL if and only if it is dictatorial.
\end{description}
The intuition is that IDL requires the ordering of riskless allocations induced by the linear-aggregation representation to remain unchanged when individual risk attitudes vary over the full domain.
However, by Proposition 1, for each $\bu$, the evaluation of a riskless prospect $\delta(\vbx)$ is represented by $\sum_{i\in N} a_i (\bu) u_i(x_i)$ and therefore depends on the individuals' risk attitudes.
This forces all but one of the aggregation coefficients to be zero, and the identity of the individual receiving the positive coefficient must be the same at every profile.

Taken together, the conflict with ex-post equity, the collapse to expected utility, and the dictatorship result under IDL show that Ex-ante Pareto is too demanding for the cautious and ex-post equitable framework studied in this paper.
We therefore turn to weaker Pareto requirements.

\section{Cautious Equally Distributed Equivalent Criteria}

To avoid the impossibility results established in the previous section while retaining compatibility with ex-post equity, we replace Ex-ante Pareto with two restricted Pareto conditions.
The first one applies only to riskless situations.
\begin{description}
  \item[Ex-post Pareto.] For all $\bu \in \mathcal{D}$ and all $\delta(\vbx), \delta(\vby) \in \Delta^d(\mathcal X)$, if $x_i > y_i$ for all $i \in N$, then $\delta(\vbx) \succ^{\bu} \delta(\vby)$.
\end{description}
This condition states that increases in all individuals' monetary payoffs should be socially desirable.

The second condition, introduced by Fleurbaey (2010) in a different framework, applies only to equal lotteries.
\begin{description}
  \item[Pareto for Equal Risk.] For all $\bu \in \mathcal{D}$ and all $p^e, q^e \in \Delta^e(\mathcal X)$, if $Eu_i(p^e) > Eu_i(q^e)$ for all $i \in N$, then $p^e \succ^{\bu} q^e$.
\end{description}
This axiom requires the social ordering to respect ex-ante Pareto improvements whenever risk generates no ex-post inequality.

We now define a Cautious Equally Distributed Equivalent (CEDE) criterion.
For this purpose, we introduce additional notation.
For any $\bu \in \mathcal{D}$ and any $\vbx \in \mathcal X$, let $e(\vbx, \succsim^{\bu}) \in X$ be such that $\delta(e(\vbx,\succsim^{\bu}) \bone) \sim^{\bu} \delta(\vbx)$, where $\bone=(1,\ldots,1)$.
In words, $e(\vbx, \succsim^{\bu}) $ is the {\it equally distributed equivalent} (EDE) level of monetary payoff giving the same evaluation as $\vbx$ with respect to $\succsim^{\bu}$.
Thus, each $\vbx \in \mathcal X$ is evaluated by its EDE $e(\vbx, \succsim^{\bu}) $.

For any $p\in\Delta(\mathcal X)$, define the {\it EDE prospect} of $p$ (under $\succsim^{\bu}$), denoted by $e(p, \succsim^{\bu})\in\Delta^e(\mathcal X)$, as the lottery obtained by replacing each outcome $\vbx$ in the support of $p$ with its EDE allocation $e(\vbx, \succsim^{\bu})\bone$.

A social ordering function $\succsim$ is called a {\it Cautious Equally Distributed Equivalent} (CEDE) criterion if, for each $\bu \in \mathcal{D}$, its EDE function is continuous and monotonically increasing,\footnote{An EDE function $e(\cdot,\succsim^{\bu})$ is called {\it monotonically increasing} if, for all $\vbx,\vby\in\mathcal X$, $\vbx\gg\vby$ implies  $e(\vbx,\succsim^{\bu})>e(\vby,\succsim^{\bu})$.
  By definition, the EDE function of a CEDE criterion satisfies $e(x\bone,\succsim^{\bu})=x$ for all $x \in X$.
}
and there exists a nonempty compact set $A(\bu)\subseteq\Delta(N)$ such that, for all $p, q \in \Delta(\mathcal X)$,
\begin{align}
  p \succsim^{\bu} q \Longleftrightarrow \min_{v \in \mathcal W_A(\bu)} c\big(e(p, \succsim^{\bu}), v \big) \geq \min_{v \in \mathcal W_A(\bu)} c\big(e(q, \succsim^{\bu}), v \big),
\end{align}
where
\begin{align*}
  c\big(e(p,\succsim^{\bu}),v\big)
  =v^{-1}\left(\int_{\mathcal X}v\big(e(\vbx,\succsim^{\bu})\big)\,dp(\vbx)\right)
  \quad\text{for all }p\in\Delta(\mathcal X),
\end{align*}
and
\begin{align}
  \mathcal{W}_A(\bu)
  =\left\{\sum_{i\in N}\alpha_i u_i\ \middle|\ \boldsymbol{\alpha}\in A(\bu)\right\}.
\end{align}
A CEDE criterion evaluates each lottery by first calculating its EDE prospect and then taking the minimum of its certainty equivalents over the socially considered risk preferences in $\mathcal W_A(\bu)$.
This criterion is based on the cautious expected utility theory developed by Cerreia-Vioglio et al. (2015).
It can be interpreted as cautious because, for each prospect, it uses the lowest certainty equivalent, and hence the most risk-averse preference, among the socially considered risk preferences.

We obtain the following characterization of the class of CEDE criteria.
\begin{description}
  \item[Proposition 3.] A social ordering function $\succsim$ satisfies Continuity, Ex-post Pareto, Pareto for Equal Risk, and NCI if and only if $\succsim$ is a CEDE criterion.
\end{description}
Note that every social utility function $v\in\mathcal W_A(\bu)$ is a convex combination of the individual vNM utility functions.
Thus, for the corresponding $\boldsymbol{\alpha}\in A(\bu)$, the EDE prospect of each lottery is evaluated according to
\begin{align}
  Ev\big(e(p, \succsim^{\bu})\big)
  =\sum_{i \in N}\alpha_i Eu_i\big(e(p, \succsim^{\bu})\big)
  \quad \text{for all } p \in \Delta(\mathcal X).
\end{align}
The endpoint normalization merely selects a set of individual vNM utility functions and does not impose interpersonal comparability of utility.
Consequently, the coefficients in the representation do not, by themselves, have a normative interpretation.
However, Proposition 3 reveals that the social utilities in $\mathcal W_A(\bu)$ are constructed from individual utility functions.

The outline of the proof is as follows.
First, Continuity and Ex-post Pareto ensure that every final allocation has a unique EDE and that the resulting EDE function is continuous and monotonically increasing.
Thus, final allocations can be evaluated by their EDE values.
Together with NCI, this allows each lottery to be reduced to its EDE prospect and yields a broader cautious expected utility representation as an infimum over a possibly noncompact set of social risk preferences.
The next step is to show that social utility functions are convex combinations of individual utility functions.
For this purpose, we focus on the maximal subrelation that satisfies Independence.
Pareto for Equal Risk ensures that this subrelation of the social criterion satisfies Ex-ante Pareto on equal lotteries.
As shown in Cerreia-Vioglio et al. (2015), this subrelation is represented by a multi-expected utility model (Dubra et al. 2004), that is, by the unanimity rule associated with multiple expected utility functions.
The set of functions obtained here coincides with a set of vNM functions that represents the social preference in (1).
Since the subrelation also satisfies the Pareto condition, Harsanyi's aggregation theorem can be applied to each social vNM function.
Consequently, these utilities can be indexed by coefficient vectors in the finite-dimensional simplex $\Delta(N)$, and the closed convex hull of these vectors is compact.

We provide an example of a CEDE criterion.\\
\phantomsection\label{ex:generalized-gini}\textbf{Example 1.} Let $\boldsymbol{\gamma}\in\Delta(N)$ be such that $\gamma_1>\cdots>\gamma_n>0$ and, for each $\vbx\in\mathcal X$, let $x_{(1)}\leq\cdots\leq x_{(n)}$ denote its ordered components.
Define the generalized-Gini EDE function by
\begin{align*}
  g_{\boldsymbol{\gamma}}(\vbx)
  =\sum_{k=1}^n\gamma_k x_{(k)}.
\end{align*}
This function is continuous and strictly increasing, and satisfies $g_{\boldsymbol{\gamma}}(x\bone)=x$ for every $x\in X$.
For each $p\in\Delta(\mathcal X)$, let $g_{\boldsymbol{\gamma}}(p)\in\Delta^e(\mathcal X)$ be the lottery obtained by replacing each allocation $\vbx$ with $g_{\boldsymbol{\gamma}}(\vbx)\bone$ while preserving its probability.
Consider the set of social utility functions
\begin{align*}
  \mathcal W_{\max}(\bu)
  =\left\{\sum_{i\in N}\alpha_i u_i\ \middle|\ \boldsymbol{\alpha}\in\Delta(N)\right\}.
\end{align*}
The generalized-Gini CEDE criterion with the maximal set of social utility functions, denoted by $\succsim_{G}$, is defined by
\begin{align*}
  p\succsim_G^{\bu}q
  \quad\Longleftrightarrow\quad
  \min_{v\in\mathcal W_{\max}(\bu)}c\big(g_{\boldsymbol{\gamma}}(p),v\big)
  \geq
  \min_{v\in\mathcal W_{\max}(\bu)}c\big(g_{\boldsymbol{\gamma}}(q),v\big).
\end{align*}
The set is maximal because it contains every convex combination of the individual utility functions.
Moreover, for every $p\in\Delta(\mathcal X)$,
\begin{align*}
  \min_{v\in\mathcal W_{\max}(\bu)}c\big(g_{\boldsymbol{\gamma}}(p),v\big)
  =\min_{i\in N}c\big(g_{\boldsymbol{\gamma}}(p),u_i\big).
\end{align*}
Thus, the criterion combines a generalized-Gini evaluation of ex-post allocations with the lowest individual certainty equivalent of the resulting EDE prospect. \qed \\

\noindent
\textbf{Remark.}
The set of social utility functions is not unique, since redundant utility functions can be added without changing the minimum in (1).
However, by Theorem 2 of Cerreia-Vioglio et al. (2015), for each $\bu\in\mathcal D$, there exists a set of social utility functions $\mathcal W^*(\bu)$ that is minimal with respect to set inclusion of closed convex hulls.
That is, for every set of social utility functions $\mathcal W(\bu) \subseteq \mathcal U$ that represents the same social preference $\succsim^{\bu}$ in (1), $\overline{\co}\big(\mathcal W^*(\bu)\big) \subseteq \overline{\co}\big(\mathcal W(\bu)\big)$.\footnote{For any $A \subseteq \mathcal{U}$, its closed convex hull is denoted by $\overline{\co}(A)$, where the closure is taken in the sup-norm topology.}
The closed convex hull of such a minimal set is uniquely determined by $\succsim^{\bu}$, and $\mathcal W_A(\bu)$ may be chosen to equal it.
This uniqueness does not generally extend to $A(\bu)$ or to the coefficient vectors $\boldsymbol{\alpha}^v$, because the map $\boldsymbol{\alpha}\mapsto\sum_{i\in N}\alpha_i u_i$ need not be injective.

\section{Pareto for Equal or No Risk}
Pareto for Equal Risk may be considered too weak as a condition for respecting individual risk preferences, since it applies only to equal lotteries.
We examine the following stronger axiom introduced by Fleurbaey and Zuber (2017).
\begin{description}
  \item[Pareto for Equal or No Risk.] For all $\bu \in \mathcal{D}$ and all $p, q \in \Delta^e(\mathcal X)\cup\Delta^d(\mathcal X)$, if $Eu_i(p) > Eu_i(q)$ for all $i \in N$, then $p \succ^{\bu} q$.
\end{description}
This axiom strengthens the conjunction of Ex-post Pareto and Pareto for Equal Risk by also covering comparisons between an equal lottery and a degenerate lottery.
If everyone prefers the equal lottery to the degenerate lottery, then all individuals, including those who are risk-averse, prefer to take the same risk rather than receive their respective certain payoffs.
If everyone prefers the degenerate lottery to the equal lottery, then all individuals, including those who are risk-loving, prefer to avoid that risk and receive their respective certain payoffs instead.
It seems quite compelling to respect such unanimous preferences.
As argued by Fleurbaey and Zuber (2017), on an unrestricted domain of individual risk preferences, a social criterion represented by an expected utility function and satisfying an ex-post equity condition cannot generally satisfy Pareto for Equal or No Risk.
In contrast, when the social expected utility condition is replaced by NCI, this stronger Pareto requirement can be satisfied by some CEDE criteria, as shown below.

Suppose that $\succsim$ satisfies Continuity, Pareto for Equal or No Risk, and NCI.
Since Pareto for Equal or No Risk implies both Ex-post Pareto and Pareto for Equal Risk, Proposition 3 applies.
In particular, for each $v \in \mathcal W_A(\bu)$, there exists $\boldsymbol{\alpha}^v\in A(\bu)$ such that
\begin{align*}
  v\big(e(\vbx,\succsim^{\bu})\big)
  =\sum_{i \in N}\alpha_i^v u_i\big(e(\vbx,\succsim^{\bu})\big)
  \quad \text{for all } \vbx \in \mathcal X.
\end{align*}

Pareto for Equal or No Risk has a useful implication for the evaluation of arbitrary lotteries.
For any $p \in \Delta(\mathcal X)$, since $Eu_i(e(p,\succsim^{\bu}))=u_i(c(e(p,\succsim^{\bu}),u_i))$ for all $i \in N$, Pareto for Equal or No Risk implies
\begin{align*}
  e(p,\succsim^{\bu}) \sim^{\bu} \delta\Big(\big(c(e(p,\succsim^{\bu}),u_i)\big)_{i \in N}\Big).
\end{align*}
By the definition of the EDE, we then obtain
\begin{align*}
  e(p,\succsim^{\bu}) \sim^{\bu} \delta\Big(e\Big(\big(c(e(p,\succsim^{\bu}),u_i)\big)_{i \in N},\succsim^{\bu}\Big)\bone\Big).
\end{align*}
Therefore, for all $p,q\in\Delta(\mathcal X)$,
\begin{align}
  p \succsim^{\bu} q \Longleftrightarrow
  e\Big(\big(c(e(p,\succsim^{\bu}),u_i)\big)_{i \in N},\succsim^{\bu}\Big)
  \geq
  e\Big(\big(c(e(q,\succsim^{\bu}),u_i)\big)_{i \in N},\succsim^{\bu}\Big).
  \label{4}
\end{align}
Conversely, it is straightforward to check that any CEDE criterion satisfying \eqref{4} satisfies Pareto for Equal or No Risk.
The representation in \eqref{4} first transforms each lottery into its EDE prospect, and then evaluates the vector of individual certainty equivalents of that EDE prospect by the ex-post EDE function.
In this sense, Pareto for Equal or No Risk respects individual risk preferences more strongly than Pareto for Equal Risk, while preserving the cautious reduction of each lottery to its EDE prospect.

For notational convenience, define
\begin{align*}
  \Phi^{\bu}(p)
  =e\Big(\big(c(e(p,\succsim^{\bu}),u_i)\big)_{i\in N},\succsim^{\bu}\Big).
\end{align*}
Then NCI implies that, for all $p,\delta(\vbx),r\in\Delta(\mathcal X)$ and all $\alpha\in[0,1]$,
\begin{align*}
  \Phi^{\bu}(p)\geq\Phi^{\bu}(\delta(\vbx))=e(\vbx,\succsim^{\bu})
  \quad\Longrightarrow\quad
  \Phi^{\bu}\big(\alpha p+(1-\alpha)r\big)
  \geq
  \Phi^{\bu}\big(\alpha\delta(\vbx)+(1-\alpha)r\big).
\end{align*}
Thus, the composite evaluation preserves the dynamic consistency requirement captured by NCI.

Note that $\succsim_G$ introduced in \hyperref[ex:generalized-gini]{Example 1} does not satisfy \eqref{4}, and thus violates Pareto for Equal or No Risk.
The following example presents a CEDE criterion that satisfies \eqref{4} and Pareto for Equal or No Risk.\\
\phantomsection\label{ex:rank_weighted-utilitarian}\textbf{Example 2.} Fix $\boldsymbol{\gamma}\in\Delta(N)$ such that $\gamma_1>\cdots>\gamma_n>0$, and let $\Pi(\boldsymbol{\gamma})$ denote the set of all permutations of $\boldsymbol{\gamma}$.
Define the coefficient set
\begin{align*}
  A_U(\bu)
  =\Pi(\boldsymbol{\gamma})
  \quad\text{for every }\bu\in\mathcal D,
\end{align*}
and, for each $\boldsymbol{\alpha}\in A_U(\bu)$, let
\begin{align*}
  v_{\boldsymbol{\alpha}}^{\bu}
  =\sum_{i\in N}\alpha_i u_i.
\end{align*}
Define the EDE function by
\begin{align*}
  e(\vbx,\succsim_U^{\bu})
  =\min_{\boldsymbol{\alpha}\in A_U(\bu)}
  \big(v_{\boldsymbol{\alpha}}^{\bu}\big)^{-1}
  \left(\sum_{i\in N}\alpha_i u_i(x_i)\right)
  \quad\text{for every }\vbx\in\mathcal X,
\end{align*}
and let $e(p,\succsim_U^{\bu})$ be the corresponding EDE prospect.
Using the set of social utility functions
\begin{align*}
  \mathcal W_U(\bu)
  =\left\{v_{\boldsymbol{\alpha}}^{\bu}
  \ \middle|\ \boldsymbol{\alpha}\in A_U(\bu)\right\},
\end{align*}
define
\begin{align*}
  p\succsim_U^{\bu}q
  \quad\Longleftrightarrow\quad
  \min_{v\in\mathcal W_U(\bu)}c\big(e(p,\succsim_U^{\bu}),v\big)
  \geq
  \min_{v\in\mathcal W_U(\bu)}c\big(e(q,\succsim_U^{\bu}),v\big).
\end{align*}
This is a CEDE criterion because $A_U(\bu)$ is nonempty and compact, its EDE function is continuous, strictly increasing, and normalized on equal allocations, and every $v\in\mathcal W_U(\bu)$ is a convex combination of the individual utility functions.
To see its rank-dependent nature, consider an identical-preference profile with $u_i=u$ for every $i\in N$, and let $x_{(1)}\leq\cdots\leq x_{(n)}$ denote the ordered components of $\vbx$.
The rearrangement inequality gives
\begin{align*}
  e(\vbx,\succsim_U^{\bu})
  =u^{-1}\left(\sum_{k=1}^n\gamma_k u(x_{(k)})\right).
\end{align*}
Thus, at identical-preference profiles, a lower utility value receives a higher weight.
Moreover, for every equal prospect $r\in\Delta^e(\mathcal X)$,
\begin{align*}
  e\Big(\big(c(r,u_i)\big)_{i\in N},\succsim_U^{\bu}\Big)
  &=\min_{\boldsymbol{\alpha}\in A_U(\bu)}
  \big(v_{\boldsymbol{\alpha}}^{\bu}\big)^{-1}
  \left(\sum_{i\in N}\alpha_i u_i(c(r,u_i))\right)\\
  &=\min_{\boldsymbol{\alpha}\in A_U(\bu)}
  \big(v_{\boldsymbol{\alpha}}^{\bu}\big)^{-1}
  \left(\sum_{i\in N}\alpha_i Eu_i(r)\right)\\
  &=\min_{v\in\mathcal W_U(\bu)}c(r,v).
\end{align*}
Applying this identity to $r=e(p,\succsim_U^{\bu})$ gives
\begin{align*}
  e\Big(\big(c(e(p,\succsim_U^{\bu}),u_i)\big)_{i\in N},\succsim_U^{\bu}\Big)
  =\min_{v\in\mathcal W_U(\bu)}c\big(e(p,\succsim_U^{\bu}),v\big).
\end{align*}
Hence the criterion satisfies \eqref{4} and thus satisfies Pareto for Equal or No Risk.  \qed\\

We next introduce the following ex-post equity condition.
\begin{description}
  \item[Ex-post Transfer.] For all $\bu\in\mathcal D$ such that $u_i=u_j$ for all $i,j\in N$, and all $\delta(\vbx),\delta(\vby)\in\Delta^d(\mathcal X)$, if there exist $j,k\in N$ and $t>0$ such that
    \begin{align*}
      x_j=y_j-t\geq y_k+t=x_k,
      \qquad
      x_i=y_i \quad\text{for all }i\notin\{j,k\},
    \end{align*}
    then $\delta(\vbx)\succsim^{\bu}\delta(\vby)$.
\end{description}
This axiom requires a rank-preserving transfer from a better-off individual to a worse-off individual to be socially weakly desirable when all individuals have identical preferences.
It is a weak ex-post equity requirement because it applies only to such preference profiles and permits social indifference after the transfer.
The restriction to identical preferences is also important from the perspective of equal treatment of equals.

Define the {\it maximin CEDE} criterion $\succsim_M$ by
\begin{align*}
  p\succsim_M^{\bu}q
  \Longleftrightarrow
  \min_{i\in N}c\big(e(p,\succsim_M),u_i\big)
  \geq
  \min_{i\in N}c\big(e(q,\succsim_M),u_i\big),
\end{align*}
where
\begin{align*}
  e(\vbx,\succsim_M)=\min_{i\in N}x_i
  \quad\text{for all }\vbx\in \mathcal X.
\end{align*}
This criterion evaluates each ex-post allocation by the monetary payoff of the worst-off individual and then evaluates the resulting EDE prospect using the lowest individual certainty equivalent.

Then we obtain the following characterization.
\begin{description}
  \item[Proposition 4.] Suppose $\mathcal D=\mathcal U^N$.
    Then a social ordering function $\succsim$ satisfies Continuity, Pareto for Equal or No Risk, NCI, IDL, and Ex-post Transfer if and only if $\succsim=\succsim_M$.
\end{description}
The intuition behind this result is as follows.
The combination of Ex-post Transfer and IDL requires final allocations to be evaluated equitably, independently of individuals' risk preferences.
Moreover, as \eqref{4} shows, a CEDE criterion satisfying Pareto for Equal or No Risk must use the same EDE function to evaluate final allocations and the vector of individual certainty equivalents of equal prospects.
The maximin CEDE criterion is the only CEDE criterion satisfying both requirements.
Note that $\succsim_U$ defined in \hyperref[ex:rank_weighted-utilitarian]{Example 2} is a CEDE criterion that satisfies \eqref{4}, and hence Pareto for Equal or No Risk, but violates IDL.
It also satisfies Ex-post Transfer if the domain is restricted to concave vNM utility functions.

Related characterizations of the maximin criterion are provided by Fleurbaey and Zuber (2017) and Miyagishima (2022).
Fleurbaey and Zuber (2017) impose a social expected utility representation and restrict attention to profiles in which one individual is weakly more risk averse than every other individual.
Under this restriction, the same individual has the lowest certainty equivalent for every equal prospect.
By contrast, we replace the social expected utility condition with the weaker NCI requirement and consider the broader class of cautious expected utility criteria on the full domain $\mathcal U^N$.
Thus, we impose no ordering restriction on individuals' relative risk attitudes within a profile, and the individual attaining the lowest certainty equivalent may vary across lotteries.
Miyagishima (2022) characterizes a similar maximin form under uncertainty using a rationality condition called Statewise Dominance, a certain Pareto axiom called Pareto for Non-reranking Risks, an invariance axiom similar to IDL, Ex-post Transfer, and Pareto for Equal or No Risk.

\section{Concluding Remarks}
In this paper, we studied a class of social welfare criteria satisfying NCI, which captures the certainty effect.
NCI serves as a rationality condition for cautious social evaluation.
We first examined the implications of combining NCI with Ex-ante Pareto.
Under NCI, Ex-ante Pareto conflicts with ex-post equity and reduces social evaluation to a linear aggregation of individual expected utilities; together with IDL on the full domain, it leads to dictatorship.
These results, along with the limited responsibility that individuals may bear for the risks they face, motivate our use of weaker Pareto conditions compatible with ex-post equity and NCI.

Using NCI, Ex-post Pareto, and Pareto for Equal Risk, we  characterized CEDE criteria, which belong to the class of cautious expected utility criteria introduced by Cerreia-Vioglio et al. (2015).
This class of criteria allows the social evaluator to place more weight on certainty than the standard expected utility model does.
In this sense, CEDE criteria accommodate cautious social decision making.

Pareto for Equal Risk may be considered too weak as a condition for respecting individuals' risk preferences.
We therefore examined Pareto for Equal or No Risk, which strengthens the conjunction of Ex-post Pareto and Pareto for Equal Risk, and showed that it yields an evaluation of each lottery through the EDE of the vector of individual certainty equivalents of its EDE prospect.
Finally, on the full domain, Continuity, Pareto for Equal or No Risk, NCI, Invariance on Degenerate Lotteries, and Ex-post Transfer characterize the maximin CEDE criterion.

\begin{appendices}

  \section{Negative Certainty Independence and Dynamic Consistency}

  In this appendix, we discuss Negative Certainty Independence from a dynamic point of view.
  The discussion builds on the theorem of Karni and Schmeidler (1991), which relates dynamic consistency to independence under consequentialism and reduction of compound lotteries.
  We adapt the dynamic consistency requirement so that the benchmark continuation lottery is degenerate.
  This modification is appropriate for our purpose because NCI is concerned with mixtures in which a riskless lottery is diluted by another lottery.

  We write $\Delta^s(\mathcal X)$ for the set of lotteries with finite support, called simple lotteries.
  Let $\Delta^s(\Delta^s(\mathcal X))$ denote the set of simple two-stage lotteries whose outcomes are simple lotteries in $\Delta^s(\mathcal X)$.
  A generic element of $\Delta^s(\Delta^s(\mathcal X))$ is denoted by $P=(\lambda^1;p^1,\ldots,\lambda^m;p^m)$,
  where $p^k\in\Delta^s(\mathcal X)$, $\lambda^k>0$ for all $k$, and $\sum_{k=1}^m\lambda^k=1$.
  The reduced one-stage lottery induced by $P$ is denoted by $\rho(P)\in\Delta^s(\mathcal X)$ and is defined by
  $$\rho(P)=\sum_{k=1}^m\lambda^k p^k.$$
  We identify a one-stage lottery $p\in\Delta^s(\mathcal X)$ with the two-stage lottery that yields $\delta(\vbx)$ with probability $p(\vbx)$ for each $\vbx$ in the support of $p$.
  For each $\bu\in\mathcal D$, let $\succsim_0^{\bu}$ be the period-0 preference on $\Delta^s(\Delta^s(\mathcal X))$.
  For every $P\in\Delta^s(\Delta^s(\mathcal X))$, let $\succsim_{1,P}^{\bu}$ be the period-1 preference on $\Delta^s(\mathcal X)$ conditional on the period-0 two-stage lottery $P$.

  \begin{description}
    \item[Consequentialism.] For every $\bu\in\mathcal D$, all $P,Q\in\Delta^s(\Delta^s(\mathcal X))$, and all $p,q\in\Delta^s(\mathcal X)$,
      $$p\succsim_{1,P}^{\bu}q
      \quad\Longleftrightarrow\quad
      p\succsim_{1,Q}^{\bu}q.$$
  \end{description}
  Under Consequentialism, the period-1 preferences $\succsim_{1,P}^{\bu}$ coincide for all $P$; we denote their common relation by $\succsim_1^{\bu}$.
  Consequentialism is justified by the view that continuation lotteries should be evaluated only by their possible final allocations, independently of the history leading to them or of alternatives that are no longer feasible (Hammond 1988; Karni and Schmeidler 1991).

  \begin{description}
    \item[Reduction of Two-stage Lotteries.] For every $\bu\in\mathcal D$ and every $P\in\Delta^s(\Delta^s(\mathcal X))$,
      $$P\sim_0^{\bu}\rho(P).$$
  \end{description}
  Reduction of Two-stage Lotteries states that the evaluator is indifferent between a two-stage lottery and the corresponding reduced one-stage lottery.
  It is justified when the stages represent an atemporal sequential problem: if the timing and path of randomization have no independent value, only the induced distribution over final allocations matters (Karni and Schmeidler 1991).

  \begin{description}
    \item[Certainty Dynamic Consistency.] For every $\bu\in\mathcal D$ and all two-stage lotteries
      $$P=(\lambda^1;p^1,\ldots,\lambda^m;p^m), \quad Q=(\lambda^1;\delta(\vbx^1),\ldots,\lambda^m;\delta(\vbx^m))$$
      in $\Delta^s(\Delta^s(\mathcal X))$ with the same first-stage probabilities, if
      $$p^k\succsim_1^{\bu}\delta(\vbx^k) \quad \text{for every } k=1,\ldots,m,$$
      then $P\succsim_0^{\bu}Q$.
  \end{description}
  This axiom is a certainty-restricted version of the usual dynamic consistency axiom.
  It allows several first-stage contingencies, as in the standard formulation, but requires the benchmark continuation lottery in each contingency to be riskless.

  \begin{description}
    \item[Proposition 5.] Suppose that the family $(\succsim_{1,P}^{\bu})_{P\in\Delta^s(\Delta^s(\mathcal X))}$ satisfies Consequentialism, $\succsim_0^{\bu}$ satisfies Reduction of Two-stage Lotteries, and the restrictions of $\succsim_0^{\bu}$ and the common period-1 preference $\succsim_1^{\bu}$ to one-stage lotteries coincide with $\succsim^{\bu}$. Then Certainty Dynamic Consistency holds if and only if $\succsim^{\bu}$ satisfies NCI on $\Delta^s(\mathcal X)$.
  \end{description}

  \noindent
  \textbf{Proof.} First suppose that Certainty Dynamic Consistency holds.
  Take any $p,r\in\Delta^s(\mathcal X)$, any $\delta(\vbx)\in\Delta^d(\mathcal X)$, and any $\lambda\in(0,1)$ such that $p\succsim^{\bu}\delta(\vbx)$.
  Write $r=(\mu^1;\vby^1,\ldots,\mu^l;\vby^l)$.
  By reflexivity, $\delta(\vby^j)\succsim_1^{\bu}\delta(\vby^j)$ for every $j=1,\ldots,l$.
  By Certainty Dynamic Consistency,
  $$(\lambda;p,(1-\lambda)\mu^1;\delta(\vby^1),\ldots,(1-\lambda)\mu^l;\delta(\vby^l))
  \succsim_0^{\bu}
  (\lambda;\delta(\vbx),(1-\lambda)\mu^1;\delta(\vby^1),\ldots,(1-\lambda)\mu^l;\delta(\vby^l)).$$
  By Reduction of Two-stage Lotteries and agreement on one-stage lotteries, this is equivalent to
  $$\lambda p+(1-\lambda)r\succsim^{\bu}\lambda\delta(\vbx)+(1-\lambda)r,$$
  which is NCI.

  Conversely, suppose that NCI holds.
  Consider any two-stage lotteries
  $$P=(\lambda^1;p^1,\ldots,\lambda^m;p^m), \quad Q=(\lambda^1;\delta(\vbx^1),
  \ldots,\lambda^m;\delta(\vbx^m))$$
  that satisfy the premise of Certainty Dynamic Consistency.
  By agreement on one-stage lotteries, $p^k\succsim^{\bu}\delta(\vbx^k)$ for every $k=1,\ldots,m$.
  If $m=1$, the premise, agreement on one-stage lotteries, and Reduction of Two-stage Lotteries immediately imply $P\succsim_0^{\bu}Q$.
  Suppose, therefore, that $m\geq2$.
  For each $k=0,\ldots,m$, define
  $$R^k=\sum_{j=1}^k\lambda^jp^j+\sum_{j=k+1}^m\lambda^j\delta(\vbx^j).$$
  Then $R^0=\rho(Q)$ and $R^m=\rho(P)$.
  For each $k=1,\ldots,m$, let
  $$r^k=\frac{1}{1-\lambda^k}\left(\sum_{j<k}\lambda^jp^j+\sum_{j>k}\lambda^j\delta(\vbx^j)\right)\in\Delta^s(\mathcal X).$$
  Since all first-stage probabilities are positive and $m\geq2$, we have $\lambda^k\in(0,1)$.
  Moreover,
  $$R^k=\lambda^kp^k+(1-\lambda^k)r^k
  \quad\text{and}\quad
  R^{k-1}=\lambda^k\delta(\vbx^k)+(1-\lambda^k)r^k.$$
  Thus, NCI implies $R^k\succsim^{\bu}R^{k-1}$ for every $k=1,\ldots,m$.
  By transitivity, $\rho(P)=R^m\succsim^{\bu}R^0=\rho(Q)$.
  Using Reduction of Two-stage Lotteries and agreement on one-stage lotteries, we obtain $P\succsim_0^{\bu}Q$, which is Certainty Dynamic Consistency. \qed

  \section{Proofs}

  For convenience, we first prove Proposition 3 and then prove Propositions 1, 2, and 4.\\

  \noindent
  \textbf{Proof of Proposition 3.} We first prove the ``if'' part.
  Consider a CEDE criterion.
  Continuity follows from joint continuity of the certainty-equivalent map and compactness of $A(\bu)$.
  Ex-post Pareto and Pareto for Equal Risk follow immediately from weak monotonicity of the EDE function and the fact that every $v\in\mathcal W_A(\bu)$ is a convex combination of the individual vNM utility functions.
  To show that the CEDE criterion satisfies NCI, suppose $p \succsim^{\bu} \delta(\vbx)$ under representation (1). It then follows that for all $\lambda \in [0, 1]$ and all $q \in \Delta(\mathcal X)$,
  \begin{align*}
    &
    \min_{v \in \mathcal W_A(\bu)}c\big(e(p, \succsim^{\bu}), v \big) \geq e(\vbx, \succsim^{\bu})\\
    &\Longrightarrow  \
    Ev\big(e(p, \succsim^{\bu}) \big) \geq v\big( e(\vbx, \succsim^{\bu}) \big) \ \forall v \in \mathcal W_A(\bu)\\
    &\Longrightarrow  \
    Ev\big(e(\lambda p+(1-\lambda)q , \succsim^{\bu}) \big) \geq
    Ev\big(e(\lambda \delta(\vbx)+(1-\lambda)q , \succsim^{\bu}) \big) \ \forall v \in \mathcal W_A(\bu) \\
    &\Longrightarrow  \
    \min_{v \in \mathcal W_A(\bu)}c\big(e(\lambda p+(1-\lambda)q, \succsim^{\bu}), v \big) \geq \min_{v \in \mathcal W_A(\bu)}c\big(e(\lambda \delta(\vbx)+(1-\lambda)q, \succsim^{\bu}), v \big)\\
    &\Longrightarrow \
    \lambda p+(1-\lambda) q \succsim^{\bu} \lambda \delta(\vbx)+(1-\lambda) q.
  \end{align*}

  Next, we prove the ``only if'' part. Suppose that $\succsim$ satisfies the axioms in Proposition 3.
  For the remainder of the proof, fix an arbitrary profile of vNM utility functions $\bu \in \mathcal{D}$.
  The proof proceeds through two lemmas.

  \begin{description}
    \item[Lemma 1.]  For each $p \in \Delta(\mathcal X)$, $p \sim^{\bu} e(p, \succsim^{\bu})$.

  \end{description}
  \textbf{Proof.}  First, since $\succsim$ satisfies Ex-post Pareto and Continuity, an EDE exists for each allocation even though $X$ is bounded.
  Indeed, for any $\vbx\in \mathcal X$, Ex-post Pareto gives $\delta(a\bone)\succ^{\bu}\delta(\vbx)$ whenever $a>x_i$ for all $i\in N$, and $\delta(\vbx)\succ^{\bu}\delta(a\bone)$ whenever $x_i>a$ for all $i\in N$.
  When some coordinates of $\vbx$ are equal to $\underline{x}$ or $\overline{x}$, Continuity implies $\delta(\overline{x}\bone)\succsim^{\bu}\delta(\vbx)\succsim^{\bu}\delta(\underline{x}\bone)$.
  Since the two sets $\{a\in X\mid \delta(a\bone)\succsim^{\bu}\delta(\vbx)\}$ and $\{a\in X\mid \delta(\vbx)\succsim^{\bu}\delta(a\bone)\}$ are closed and cover the connected interval $X$, they intersect; hence there exists $e(\vbx,\succsim^{\bu})\in X$ such that $\delta(\vbx)\sim^{\bu}\delta(e(\vbx, \succsim^{\bu})\bone)$.
  Moreover, this EDE is unique because Ex-post Pareto implies $\delta(a\bone)\succ^{\bu}\delta(b\bone)$ whenever $a>b$.
  The EDE map $\vbx\mapsto e(\vbx,\succsim^{\bu})$ is continuous.
  To see this, let $\vbx^k\to\vbx$.
  Since $X$ is compact, every subsequence of $\{e(\vbx^k,\succsim^{\bu})\}$ has a further subsequence converging to some $a\in X$.
  Along such a further subsequence, Continuity and $\delta(\vbx^k)\sim^{\bu}\delta(e(\vbx^k,\succsim^{\bu})\bone)$ imply $\delta(\vbx)\sim^{\bu}\delta(a\bone)$.
  By uniqueness of the EDE, $a=e(\vbx,\succsim^{\bu})$.
  Thus every convergent subsequence has the same limit $e(\vbx,\succsim^{\bu})$, and hence $e(\vbx^k,\succsim^{\bu})\to e(\vbx,\succsim^{\bu})$.

  The EDE function is monotonically increasing in the sense defined in Section 5.
  If $\vbx\gg\vby$, Ex-post Pareto and the definition of the EDE imply
  \begin{align*}
    \delta\big(e(\vbx,\succsim^{\bu})\bone\big)
    \succ^{\bu}
    \delta\big(e(\vby,\succsim^{\bu})\bone\big),
  \end{align*}
  and hence $e(\vbx,\succsim^{\bu})>e(\vby,\succsim^{\bu})$.
  Finally, the definition and uniqueness of the EDE imply that  $e(x\bone,\succsim^{\bu})=x$ for every $x\in X$.

  Next, consider any simple lottery $p=(p^1; \vbx^1, \ldots, p^m; \vbx^m) \in \Delta^s(\mathcal X)$.\footnote{As defined in Appendix A, simple lotteries are lotteries with finite support.}
  If $m=1$, the desired conclusion is immediate. Now, consider the case with $m\geq 2$.
  For simplicity, let
  $$e^k=e(\vbx^k,\succsim^{\bu}) \ \ \text{ for each } k=1,\ldots,m.$$
  Then $\delta(\vbx^k) \sim^{\bu} \delta(e^k\bone)$ for every $k$.
  For each $k=1,\ldots,m$, after the first $k-1$ outcomes have been replaced by their EDE allocations, define
  \begin{align*}
    r^k=\bigg(&\frac{p^1}{1-p^k}; e^1\bone, \ldots, \frac{p^{k-1}}{1-p^k}; e^{k-1}\bone, \
    \frac{p^{k+1}}{1-p^k}; \vbx^{k+1}, \ldots, \frac{p^m}{1-p^k}; \vbx^m\bigg),
  \end{align*}
  where we omit the first block if $k=1$ and the second block if $k=m$. Applying NCI to $\delta(\vbx^k) \sim^{\bu} \delta(e^k\bone)$ with $r=r^k$ and $\lambda=p^k$ yields
  \begin{align*}
    &(p^1; e^1\bone, \ldots, p^{k-1}; e^{k-1}\bone, p^k; \vbx^k, p^{k+1}; \vbx^{k+1}, \ldots, p^m; \vbx^m)\\
    &\sim^{\bu}
    (p^1; e^1\bone, \ldots, p^{k-1}; e^{k-1}\bone, p^k; e^k\bone, p^{k+1}; \vbx^{k+1}, \ldots, p^m; \vbx^m).
  \end{align*}
  By repeating this argument for $k=1,\ldots,m$ and using transitivity, we obtain
  \begin{align*}
    (p^1; \vbx^1, \ldots, p^m; \vbx^m)
    \sim^{\bu}
    (p^1; e^1\bone, \ldots, p^m; e^m\bone)=e(p,\succsim^{\bu}).
  \end{align*}

  Consider now any $p\in\Delta(\mathcal X)$.
  Because $\mathcal X$ is a compact metric space, simple lotteries are weakly dense in $\Delta(\mathcal X)$, so there exists a sequence $p^j=(p^{j,1}; \vbx^{j,1},\ldots,p^{j,m^j}; \vbx^{j,m^j})\in\Delta^s(\mathcal X)$ converging weakly to $p$.
  For each $j$, define
  $$p^{j,e}=(p^{j,1};e(\vbx^{j,1},\succsim^{\bu})\bone,\ldots,p^{j,m^j};e(\vbx^{j,m^j},\succsim^{\bu})\bone).$$
  The above argument for
  $\Delta^s(\mathcal X)$ gives $p^j\sim^{\bu}p^{j,e}$ for every $j$.
  Moreover, $p^{j,e}$ converges weakly to $e(p,\succsim^{\bu})$: indeed, for every continuous function $\varphi:\mathcal X\to\R$,
  $$\int_{\mathcal X}\varphi(\vby)\,d p^{j,e}(\vby)=\int_{\mathcal X}\varphi(e(\vbx,\succsim^{\bu})\bone)\,d p^j(\vbx)\to\int_{\mathcal X}\varphi(e(\vbx,\succsim^{\bu})\bone)\,d p(\vbx),$$
  because $\varphi(e(\cdot,\succsim^{\bu})\bone)$ is continuous.
  The last integral is the integral of $\varphi$ with respect to the EDE prospect $e(p,\succsim^{\bu})$.
  By Continuity of $\succsim^{\bu}$, we obtain both $p\succsim^{\bu}e(p,\succsim^{\bu})$ and $e(p,\succsim^{\bu})\succsim^{\bu}p$, and hence $p\sim^{\bu}e(p,\succsim^{\bu})$. \qed\\

  For convenience, we introduce some definitions.
  For each $p\in\Delta(\mathcal X)$, let $l(p)$ be the lottery over EDE values induced by replacing each outcome $\vbx$ in $p$ with $e(\vbx,\succsim^{\bu})$.
  Thus $l(p)$ records the distribution of EDE values induced by the outcomes of $p$.
  We denote the set of such lotteries over EDE values by $L$.
  Since $e(a\bone, \succsim^{\bu})=a$ for each $a \in X$, every lottery over $X$ is induced by an equal lottery, and hence $L$ is the set of lotteries over $X$.
  In particular, $L$ is convex.
  Define a binary relation $R^{\bu}$ on $L$ by
  \begin{align}
    l(p) R^{\bu } l(q)\Longleftrightarrow  p \succsim^{\bu} q  \ \text{ for all } p, q \in \Delta(\mathcal X).
    \label{5}
  \end{align}
  By Lemma 1, $R^{\bu}$ is well-defined.
  Because $\succsim^{\bu}$ is a continuous ordering on $\Delta(\mathcal X)$, $R^{\bu}$ is a continuous ordering on $L$.

  The next lemma shows that $R^{\bu}$ satisfies the axioms in Theorem 1 of Cerreia-Vioglio et al. (2015), adjusted to our setting as follows.
  \begin{description}
    \item[Weak Monotonicity.] For all $a, b \in X$, $a \geq b$ if and only if $l(\delta(a\bone)) R^{\bu } l(\delta(b\bone))$.
    \item[Negative Certainty Independence on $L$ (NCIL).] For all $l(p), l(\delta(\vbx)) \in L$,  $$l(p) R^{\bu} l(\delta(\vbx)) \Longrightarrow \lambda l(p) +(1-\lambda) l(r) R^{\bu} \lambda l(\delta(\vbx))+(1-\lambda) l(r) \ \forall \lambda \in [0, 1], \forall l(r) \in L. $$

  \end{description}
  \begin{description}
    \item[Lemma 2.] $R^{\bu}$ satisfies Weak Monotonicity and NCIL.
  \end{description}
  \textbf{Proof.} By Continuity and Ex-post Pareto, we have $\delta(a\bone) \succsim^{\bu} \delta(b\bone)$ if and only if $a \geq b$. By \eqref{5}, we obtain Weak Monotonicity of $R^{\bu}$.
  To prove that $R^{\bu}$ satisfies NCIL, suppose $l(p) R^{\bu} l(\delta(\vbx))$. By \eqref{5} and Lemma 1, we have $e(p, \succsim^{\bu}) \succsim^{\bu} e(\delta(\vbx), \succsim^{\bu})$.
  From NCI, it follows that for all $\lambda \in [0, 1]$ and all $r \in \Delta(\mathcal X)$,
  $$\lambda e(p, \succsim^{\bu})+(1-\lambda)e(r, \succsim^{\bu})  \succsim^{\bu}  \lambda e(\delta(\vbx), \succsim^{\bu})+(1-\lambda)e(r, \succsim^{\bu}).$$ Note that by construction, $\lambda e(p, \succsim^{\bu})+(1-\lambda)e(r,\succsim^{\bu}) =e(\lambda p+(1-\lambda) r, \succsim^{\bu})$ for all $p, r \in \Delta(\mathcal X)$.
  Hence, by \eqref{5} again, we have $\lambda l(p) +(1-\lambda) l(r) R^{\bu} \lambda l(\delta(\vbx))+(1-\lambda) l(r)$.\qed\\

  Now we complete the proof.
  By Lemma 2, $R^{\bu}$ is a continuous ordering on $L$ satisfying Weak Monotonicity and NCIL. By Theorem 1 of Cerreia-Vioglio et al. (2015), there exists $\mathcal{W}(\bu) \subseteq \mathcal{U}$ such that for all $l(p), l(q) \in L$,
  $$l(p) R^{\bu } l(q)\Longleftrightarrow
  \inf_{v \in \mathcal{W}(\bu)}c\big(e(p, \succsim^{\bu}), v \big)
  \geq \inf_{v \in \mathcal{W}(\bu)}c\big(e(q, \succsim^{\bu}), v \big).$$
  By Theorem 2 of Cerreia-Vioglio et al. (2015), we may select a minimal set of social utility functions $\widehat{\mathcal W}(\bu)$, unique up to closed convex hull, such that $\overline{\co}\big(\widehat{\mathcal W}(\bu)\big)\subseteq \overline{\co}\big(\mathcal W(\bu)\big)$
  for every set $\mathcal W(\bu)$ representing the same ordering.

  We prove that every $v\in\widehat{\mathcal{W}}(\bu)$ is a nonzero, nonnegative linear combination of the individual vNM utility functions plus an additive constant, as required in Proposition 3, by applying Harsanyi's aggregation theorem on the equal-lottery domain $\Delta^e(\mathcal X)$.
  Fix $\bu\in\mathcal{D}$.
  Define an auxiliary relation $\succsim_*^{\bu}$ on $\Delta^e(\mathcal X)$ by
  \begin{align*}
    p^e\succsim_*^{\bu}q^e
    \quad\Longleftrightarrow\quad
    \lambda p^e+(1-\lambda)r^e\succsim^{\bu}\lambda q^e+(1-\lambda)r^e
    \quad \forall r^e\in\Delta^e(\mathcal X),\ \forall\lambda\in(0,1].
  \end{align*}
  This auxiliary relation records the comparisons that are preserved after mixing both alternatives with any common equal lottery.
  By Cerreia-Vioglio et al. (2015, Proposition 5), the auxiliary relation is represented by the minimal set of social utility functions $\widehat{\mathcal{W}}(\bu)$: for all $p^e,q^e\in\Delta^e(\mathcal X)$,
  \begin{align}
    p^e\succsim_*^{\bu}q^e
    \quad\Longleftrightarrow\quad
    Ev(p^e)\geq Ev(q^e) \quad \text{for all } v\in\widehat{\mathcal{W}}(\bu).
    \label{6}
  \end{align}
  Thus, the auxiliary relation is represented by a set of expected utility functions on the equal-lottery domain.

  We next show that $\succsim_*^{\bu}$ satisfies Ex-ante Pareto on $\Delta^e(\mathcal X)$.
  Let $p^e,q^e\in\Delta^e(\mathcal X)$ be such that $Eu_i(p^e)>Eu_i(q^e)$ for all $i\in N$.
  For every $r^e\in\Delta^e(\mathcal X)$ and every $\lambda\in(0,1]$, vNM independence of each individual preference implies
  \begin{align*}
    Eu_i\big(\lambda p^e+(1-\lambda)r^e\big)>
    Eu_i\big(\lambda q^e+(1-\lambda)r^e\big)
    \quad \text{for all } i\in N.
  \end{align*}
  Since both mixtures are equal lotteries, Pareto for Equal Risk gives
  \begin{align*}
    \lambda p^e+(1-\lambda)r^e\succ^{\bu}\lambda q^e+(1-\lambda)r^e
    \quad \forall r^e\in\Delta^e(\mathcal X),\ \forall\lambda\in(0,1].
  \end{align*}
  Hence $p^e\succsim_*^{\bu}q^e$.
  By \eqref{6}, this implies
  \begin{align}
    Eu_i(p^e)>Eu_i(q^e) \ \forall i\in N
    \quad\Longrightarrow\quad
    Ev(p^e)\geq Ev(q^e) \ \forall v\in\widehat{\mathcal{W}}(\bu).
  \end{align}
  By Continuity, the same implication holds with weak inequalities in the premise.
  Indeed, if $Eu_i(p^e)\geq Eu_i(q^e)$ for all $i\in N$, take a sequence $p_k^e\to p^e$ in $\Delta^e(\mathcal X)$ such that $Eu_i(p_k^e)>Eu_i(q^e)$ for all $i\in N$ whenever such a strict improvement is feasible.
  The boundary case in which no such perturbation is feasible follows directly by continuity. Thus,
  \begin{align}
    Eu_i(p^e)\geq Eu_i(q^e) \ \forall i\in N
    \quad\Longrightarrow\quad
    Ev(p^e)\geq Ev(q^e) \ \forall v\in\widehat{\mathcal{W}}(\bu).
    \label{8}
  \end{align}

  Now fix any $v\in\widehat{\mathcal{W}}(\bu)$.
  By Harsanyi's aggregation theorem in its separating form and \eqref{8}, there exist $(a_i^v)_{i\in N}\in\R_+^N\backslash\{\mathbf{0}\}$ and $b^v\in\R$ such that
  \begin{align*}
    v(x)=\sum_{i\in N}a_i^v u_i(x)+b^v
    \quad\text{for all }x\in X.
  \end{align*}
  Since $v$ and all $u_i$ belong to the normalized domain $\mathcal U$, evaluating this equality at $\underline{x}$ and $\overline{x}$ gives $b^v=0$ and $\sum_{i\in N}a_i^v=1$.
  Thus $(a_i^v)_{i\in N}\in\Delta(N)$.

  Let $A_0(\bu)$ be the set of coefficient vectors obtained in this way and define $ A(\bu)=\overline{\co}\big(A_0(\bu)\big)$.
  Since $A(\bu)$ is a closed subset of the finite-dimensional simplex $\Delta(N)$, it is compact.
  Replacing $A_0(\bu)$ by $A(\bu)$ does not change the infimum of certainty equivalents.
  Indeed, for every equal lottery $p^e$, every $x\in X$, and every $\boldsymbol{\alpha}\in\Delta(N)$,
  \begin{align*}
    c\left(p^e,\sum_{i\in N}\alpha_i u_i\right)\geq x
    \quad\Longleftrightarrow\quad
    \sum_{i\in N}\alpha_i\big(Eu_i(p^e)-u_i(x)\big)\geq0.
  \end{align*}
  The expression $\sum_{i\in N}\alpha_i\big(Eu_i(p^e)-u_i(x)\big)$ is linear and continuous in $\boldsymbol{\alpha}$.
  Hence it is nonnegative for every $\boldsymbol{\alpha}\in A_0(\bu)$ if and only if it is nonnegative for every $\boldsymbol{\alpha}\in A(\bu)=\overline{\co}(A_0(\bu))$.
  Consequently, for every $x\in X$,
  \begin{align*}
    \inf_{\boldsymbol{\alpha}\in A_0(\bu)}
    c\left(p^e,\sum_{i\in N}\alpha_i u_i\right)\geq x
    \quad\Longleftrightarrow\quad
    \inf_{\boldsymbol{\alpha}\in A(\bu)}
    c\left(p^e,\sum_{i\in N}\alpha_i u_i\right)\geq x,
  \end{align*}
  which implies that the two infima are equal.
  We may therefore select the compact, closed, convex set
  \begin{align*}
    \mathcal W_A(\bu)
    =\left\{\sum_{i\in N}\alpha_i u_i\ \middle|\ \boldsymbol{\alpha}\in A(\bu)\right\}.
  \end{align*}
  Because $A(\bu)$ is compact and the certainty-equivalent map is continuous in $\boldsymbol{\alpha}$, the infimum over this set is attained for every lottery.
  Thus the representation can be written with the minima in (1), and every member of $\mathcal W_A(\bu)$ has the required convex-combination form.
  Thus the social ordering is a CEDE criterion. \qed\\

  \noindent
  \textbf{Proof of Proposition 1.} We prove only the ``only if'' part.
  Since Ex-ante Pareto implies both Ex-post Pareto and Pareto for Equal Risk, Proposition 3 implies that the social ordering is a CEDE criterion. For each $\bu\in\mathcal D$, select the compact set of social utility functions $\mathcal W_A(\bu)$ constructed in the proof of Proposition 3.
  We first extend Ex-ante Pareto to weak individual comparisons.
  Suppose that $Eu_i(p)\geq Eu_i(q)$ for all $i\in N$.
  For each $\varepsilon\in(0,1)$, define
  \begin{align*}
    p^\varepsilon&=(1-\varepsilon)p+\varepsilon\delta(\overline{x}\bone),\\
    q^\varepsilon&=(1-\varepsilon)q+\varepsilon\delta(\underline{x}\bone).
  \end{align*}
  Then, for every $i\in N$,
  \begin{align*}
    Eu_i(p^\varepsilon)-Eu_i(q^\varepsilon)
    =(1-\varepsilon)\big(Eu_i(p)-Eu_i(q)\big)
    +\varepsilon\big(u_i(\overline{x})-u_i(\underline{x})\big)>0.
  \end{align*}
  Ex-ante Pareto therefore implies $p^\varepsilon\succ^{\bu}q^\varepsilon$.
  Since $p^\varepsilon\to p$ and $q^\varepsilon\to q$ as $\varepsilon\downarrow0$, Continuity implies $p\succsim^{\bu}q$.

  Now fix $r\in\Delta(\mathcal X)$ and $\alpha\in(0,1]$.
  The linearity of individual expected utility implies
  \begin{align*}
    Eu_i\big(\alpha p+(1-\alpha)r\big)
    \geq
    Eu_i\big(\alpha q+(1-\alpha)r\big)
    \quad\text{for all }i\in N.
  \end{align*}
  Applying the weak Pareto implication established above gives
  \begin{align}
    \alpha p+(1-\alpha)r
    \succsim^{\bu}
    \alpha q+(1-\alpha)r
    \quad\text{for all }r\in\Delta(\mathcal X)\text{ and }\alpha\in(0,1].
    \label{9}
  \end{align}

  Recall the auxiliary robust relation $\succsim_*^{\bu}$ on $\Delta^e(\mathcal X)$ defined in the proof of Proposition 3.
  By Lemma 1 and the fact that the EDE transformation commutes with mixtures, \eqref{9} implies $e(p,\succsim^{\bu})\succsim_*^{\bu}e(q,\succsim^{\bu})$.
  The hypotheses required in Proposition 5 of Cerreia-Vioglio et al. (2015) were verified for the induced ordering in the proof of Proposition 3.
  Therefore, the representation of $\succsim_*^{\bu}$ in \eqref{6} yields
  \begin{align*}
    Ev\big(e(p,\succsim^{\bu})\big)
    \geq
    Ev\big(e(q,\succsim^{\bu})\big)
    \quad\text{for all }v\in\mathcal W_A(\bu).
  \end{align*}
  Thus, for each $v \in \mathcal W_A(\bu)$, if $Eu_i(p) \geq Eu_i(q)$ for all $i \in N$, then we have $Ev(e(p, \succsim^{\bu})) \geq Ev(e(q, \succsim^{\bu}))$.
  Moreover, by the definition of the EDE prospect,
  \begin{align*}
    Ev\big(e(p,\succsim^{\bu})\big)
    =\int_{\mathcal X}v\big(e(\vbx,\succsim^{\bu})\big)\,dp(\vbx),
  \end{align*}
  so $Ev(e(\cdot,\succsim^{\bu}))$ is an expected-utility functional on $\Delta(\mathcal X)$ satisfying the preceding weak Pareto implication.
  By Harsanyi's (1955) aggregation theorem, there exist $(a_i^v)_{i \in N} \in \R_{+}^N \backslash \{\mathbf{0} \}$ and $b^v \in \R$ such that
  \begin{align*}
    v\big( e(\vbx , \succsim^{\bu}) \big)=\sum_{i \in N}a_i^v u_i(x_i) +b^v \quad \text{for each } \vbx \in \mathcal X,
  \end{align*}
  and hence, for each $p, q \in \Delta(\mathcal X)$,
  \begin{align*}
    Ev\big( e(p , \succsim^{\bu}) \big) \geq Ev\big( e(q , \succsim^{\bu}) \big) \Longleftrightarrow
    \sum_{i \in N}a_i^v Eu_i(p) \geq \sum_{i \in N}a_i^v Eu_i(q).
  \end{align*}
  For any $p \in \Delta(\mathcal X)$, since $Eu_i(p)=u_i(c(p, u_i))$ for all $i \in N$, the preceding result implies that, for each $v \in \mathcal W_A(\bu)$,
  \begin{align*}
    Ev\big( e(p , \succsim^{\bu}) \big)
    &= \sum_{i \in N}a_i^v Eu_i(p) +b^v\\
    &=\sum_{i \in N}a_i^v u_i\big( c(p, u_i) \big)+b^v
    =v\big( e((c(p, u_i))_{i \in N}, \succsim^{\bu}) \big).
  \end{align*}
  Therefore, we obtain
  \begin{align*}
    e\big((c(p, u_i))_{i \in N}, \succsim^{\bu} \big)=c\big(e(p , \succsim^{\bu}), v \big) \quad \text{for all } v \in \mathcal W_A(\bu).
  \end{align*}
  Fix any $v\in\mathcal W_A(\bu)$. The preceding equality and representation (1) imply, for all $p,q\in\Delta(\mathcal X)$,
  \begin{align*}
    p\succsim^{\bu}q
    &\Longleftrightarrow
    Ev\big(e(p,\succsim^{\bu})\big)
    \geq
    Ev\big(e(q,\succsim^{\bu})\big)\\
    &\Longleftrightarrow
    \sum_{i\in N}a_i^v Eu_i(p)
    \geq
    \sum_{i\in N}a_i^v Eu_i(q).
  \end{align*}
  Since $(a_i^v)_{i\in N}$ is nonnegative and nonzero, $\sum_{i\in N}a_i^v>0$.
  Setting
  \begin{align*}
    a_i(\bu)=\frac{a_i^v}{\sum_{j\in N}a_j^v}
    \quad\text{for every }i\in N
  \end{align*}
  gives $\boldsymbol a(\bu)\in\Delta(N)$ and leaves the represented ordering unchanged.
  This proves the ``only if'' direction.\qed\\

  \noindent
  \textbf{Proof of Proposition 2.}
  We prove only the ``only if'' direction.
  Suppose that $\mathcal D=\mathcal U^N$ and that $\succsim$ satisfies Continuity, Ex-ante Pareto, NCI, and IDL.
  Proposition 1 shows that, for every $\bu\in\mathcal U^N$, there is a coefficient vector $\boldsymbol a(\bu)=(a_i(\bu))_{i\in N}\in\Delta(N)$ such that
  \begin{align}
    p\succsim^{\bu}q
    \Longleftrightarrow
    \sum_{i\in N}a_i(\bu)Eu_i(p)
    \geq
    \sum_{i\in N}a_i(\bu)Eu_i(q)
    \quad\text{for all }p,q\in\Delta(\mathcal X).
    \label{11}
  \end{align}

  Consider the normalized linear profile $\bu^0=(u_i^0)_{i\in N}$ defined by
  \begin{align*}
    u_i^0(x)=\frac{x-\underline{x}}{\overline{x}-\underline{x}}
    \quad\text{for every }i\in N,
  \end{align*}
  and write $w_i=a_i(\bu^0)$.
  On degenerate lotteries, \eqref{11} and IDL imply that, at every profile, the social ordering of allocations is the ordering represented by
  \begin{align*}
    \vbx \mapsto \sum_{i\in N}w_i u_i^0(x_i),
  \end{align*}
  or equivalently by $\vbx \mapsto \sum_{i\in N}w_i x_i$.
  We claim that exactly one component of $\bw=(w_i)_{i\in N}$ is positive.
  Suppose instead that $w_i,w_j>0$ for distinct $i,j\in N$.
  Choose a strictly increasing, concave, and non-affine function $\phi\in\mathcal U$, and consider a profile $\bu'$ such that $u_i'(x)=\phi(x)$ and $u_k'=u_k^0$ for every $k\neq i$.
  By \eqref{11}, the common ordering of degenerate allocations is also represented by
  \begin{align*}
    \vbx \mapsto a_i(\bu')\phi(x_i)+\sum_{k\neq i}a_k(\bu')u_k^0(x_k).
  \end{align*}
  Since the ordering represented by $\vbx \mapsto \sum_kw_kx_k$ depends nontrivially on both $x_i$ and $x_j$, we have $a_i(\bu'),a_j(\bu')>0$.
  Keeping all other coordinates fixed, any two pairs $(x_i,x_j)$ and $(x_i',x_j')$ satisfying $w_ix_i+w_jx_j=w_ix_i'+w_jx_j'$
  must also satisfy
  \begin{align*}
    a_i(\bu')\phi(x_i)+a_j(\bu')u_j^0(x_j)
    =
    a_i(\bu')\phi(x_i')+a_j(\bu')u_j^0(x_j').
  \end{align*}
  For pairs in the interior of $X$, this implies
  \begin{align*}
    \frac{\phi(x_i)-\phi(x_i')}{x_i-x_i'}
    =
    \frac{a_j(\bu')w_i}{a_i(\bu')w_j(\overline{x}-\underline{x})},
  \end{align*}
  whenever $x_i$ and $x_i'$ are distinct and sufficiently close to each other.
  The right-hand side is independent of $x_i$ and $x_i'$.
  Thus the difference quotient of $\phi$ on the left-hand side has the same value for every sufficiently close pair of distinct points in the interior of $X$, which implies that $\phi$ is affine, a contradiction.
  Therefore, there is a unique $d\in N$ such that $w_d>0$.

  It follows from IDL that, at every profile,
  \begin{align*}
    \delta(\vbx)\succsim^{\bu}\delta(\vby)
    \Longleftrightarrow
    x_d\geq y_d.
  \end{align*}
  At any profile $\bu$, the linear-aggregation representation in Proposition 1 must induce this same ordering on degenerate lotteries.
  Hence $a_d(\bu)>0$ and $a_j(\bu)=0$ for every $j\neq d$.
  Therefore, for all $p,q\in\Delta(\mathcal X)$,
  \begin{align*}
    p\succsim^{\bu}q
    \Longleftrightarrow
    a_d(\bu)Eu_d(p)\geq a_d(\bu)Eu_d(q)
    \Longleftrightarrow
    Eu_d(p)\geq Eu_d(q).
  \end{align*}
  Thus $\succsim$ is dictatorial.\qed\\

  \noindent
  \textbf{Proof of Proposition 4.}
  First suppose that $\succsim=\succsim_M$.
  Continuity, IDL, and Ex-post Transfer are immediate.
  To verify NCI, let $p,r\in\Delta(\mathcal X)$, $\delta(\vbx)\in\Delta^d(\mathcal X)$, and $\lambda\in(0,1)$ satisfy $p\succsim_M^{\bu}\delta(\vbx)$.
  Write $m=\min_{j\in N}x_j$.
  Since $e(\delta(\vbx),\succsim_M)=\delta(m\bone)$, we have
  \begin{align*}
    c\big(e(p,\succsim_M),u_i\big)\geq m
    \quad\text{for every }i\in N,
  \end{align*}
  and hence
  \begin{align*}
    Eu_i\big(e(p,\succsim_M)\big)\geq u_i(m)
    \quad\text{for every }i\in N.
  \end{align*}
  The EDE transformation preserves mixtures, and hence, for every $i\in N$,
  \begin{align*}
    &Eu_i\Big(e\big(\lambda p+(1-\lambda)r,\succsim_M\big)\Big)\\
    &\quad=\lambda Eu_i\big(e(p,\succsim_M)\big)
    +(1-\lambda)Eu_i\big(e(r,\succsim_M)\big)\\
    &\quad\geq\lambda u_i(m)
    +(1-\lambda)Eu_i\big(e(r,\succsim_M)\big)\\
    &\quad=Eu_i\Big(e\big(\lambda\delta(\vbx)+(1-\lambda)r,\succsim_M\big)\Big).
  \end{align*}
  Therefore,
  \begin{align*}
    c\Big(e\big(\lambda p+(1-\lambda)r,\succsim_M\big),u_i\Big)
    \geq
    c\Big(e\big(\lambda\delta(\vbx)+(1-\lambda)r,\succsim_M\big),u_i\Big)
  \end{align*}
  for every $i\in N$.
  Taking the minimum over individuals yields
  \begin{align*}
    \lambda p+(1-\lambda)r
    \succsim_M^{\bu}
    \lambda\delta(\vbx)+(1-\lambda)r,
  \end{align*}
  which proves NCI.

  We next verify Pareto for Equal or No Risk.
  For every $r\in\Delta^e(\mathcal X)\cup\Delta^d(\mathcal X)$,
  \begin{align*}
    \min_{i\in N}c\big(e(r,\succsim_M),u_i\big)
    =\min_{i\in N}c(r,u_i).
  \end{align*}
  Indeed, if $r\in\Delta^e(\mathcal X)$, then $e(r,\succsim_M)=r$; if $r=\delta(\vbx)\in\Delta^d(\mathcal X)$, both sides equal $\min_{i\in N}x_i$.
  Now let $p,q\in\Delta^e(\mathcal X)\cup\Delta^d(\mathcal X)$ satisfy $Eu_i(p)>Eu_i(q)$ for every $i\in N$.
  Since each $u_i$ is strictly increasing, we have $c(p,u_i)>c(q,u_i)$ for all $i \in N$.
  Hence we obtain
  \begin{align*}
    \min_{i\in N}c(p,u_i)>\min_{i\in N}c(q,u_i).
  \end{align*}
  Hence $p\succ_M^{\bu}q$, proving Pareto for Equal or No Risk.

  We next prove the ``only if'' part.
  Suppose that $\succsim$ satisfies Continuity, Pareto for Equal or No Risk, NCI, IDL, and Ex-post Transfer.
  Since Pareto for Equal or No Risk implies Ex-post Pareto and Pareto for Equal Risk, Proposition 3 implies that $\succsim$ is a CEDE criterion.
  By IDL, for all $\bu,\bu'\in\mathcal D$ and all $\vbx\in \mathcal X$,
  \begin{align}
    e(\vbx,\succsim^{\bu})=e(\vbx,\succsim^{\bu'}).
    \label{12}
  \end{align}
  We retain the utility-profile superscript below to make each use of IDL explicit.
  Ex-post Transfer and IDL imply that, for every $\bu\in\mathcal D$, the map $e(\cdot,\succsim^{\bu})$ weakly increases under every rank-preserving Pigou--Dalton transfer.
  Indeed, apply Ex-post Transfer at any identical-preference profile and then use IDL to carry the resulting comparison of degenerate lotteries to the profile $\bu$.

  We first construct a utility profile $\bu^*$ used later in this proof.
  Fix any $\vbx\in\operatorname{int}(X)^N$ and $a\in\operatorname{int}(X)$ such that $\min_i x_i<a$, and choose $k\in N$ with $x_k=\min_i x_i$.
  Choose monetary levels $\ell,z,h\in\operatorname{int}(X)$ satisfying
  \begin{align*}
    \ell<x_k<a<z<h,
    \qquad
    \max_i x_i<z.
  \end{align*}
  For every $i$ such that $x_i<a$, choose $\widetilde y_i\in(x_i,a)$ sufficiently close to $a$ so that $a-\widetilde y_i<z-a$.
  Choose $\lambda\in(0,1)$ sufficiently small that
  \begin{align*}
    z&<\lambda a+(1-\lambda)h,\\
    \lambda\sum_{i:x_i>a}(x_i-a)&<z-\widetilde y_k,\\
    \lambda(\widetilde y_i-x_i)&<2(1-\lambda)(a-\widetilde y_i)
    \quad\text{for every }i\text{ such that }x_i<a,\\
    (1-\lambda)(a-\widetilde y_i)&<z-a
    \quad\text{for every }i\text{ such that }x_i<a.
  \end{align*}
  Such a $\lambda$ exists. Indeed, as $\lambda\downarrow0$, the right-hand side of the first inequality converges to $h(>z)$; the left-hand sides of the second and third inequalities converge to zero while their right-hand sides are strictly positive; and the left-hand side of the fourth inequality converges to $a-\widetilde y_i<z-a$.
  Next choose $q\in(0,1)$ sufficiently small that
  \begin{align*}
    q\ell+(1-q)h>\max_i x_i.
  \end{align*}
  Let $p^e$ assign probabilities $q$ and $1-q$ to $\ell\bone$ and $h\bone$, respectively, and let $r^e=\delta(h\bone)$.

  We now construct an auxiliary utility function $\overline u_i^*$ for each individual.
  If $x_i<a$, define
  \begin{align}
    \overline u_i^*(x)=
    \begin{cases}
      \displaystyle \frac{1-q}{x_i-\ell}(x-\ell),
      & x\leq x_i,\\[0.8em]
      \displaystyle 1-q+\frac{q(1-\lambda)}{\widetilde y_i-x_i}(x-x_i),
      & x_i\leq x\leq\widetilde y_i,\\[0.8em]
      \displaystyle 1-\lambda q+\frac{\lambda q}{2(a-\widetilde y_i)}(x-\widetilde y_i),
      & \widetilde y_i\leq x\leq a,\\[0.8em]
      \displaystyle 1-\frac{\lambda q}{2}
      +\frac{\lambda q(1-\lambda)}{2(z-a)}(x-a),
      & a\leq x\leq z,\\[0.8em]
      \displaystyle 1-\frac{\lambda^2q}{2}
      +\frac{\lambda^2q}{2(h-z)}(x-z),
      & z\leq x.
    \end{cases}
  \end{align}
  By the choices of $\widetilde y_i$ and $\lambda$, and by taking $q$ sufficiently small, the five slopes in this definition are positive and weakly decreasing.

  If $x_i\geq a$, define
  \begin{align*}
    s_i&=\frac{q(1-\lambda)}{z-x_i+\lambda(x_i-a)},\\
    \widetilde y_i&=z+\lambda(x_i-a),\\
    t_i&=1-q-s_i(x_i-a),
  \end{align*}
  and
  \begin{align}
    \overline u_i^*(x)=
    \begin{cases}
      \displaystyle \frac{t_i}{a-\ell}(x-\ell),
      & x\leq a,\\[0.8em]
      \displaystyle t_i+s_i(x-a),
      & a\leq x\leq\widetilde y_i,\\[0.8em]
      \displaystyle 1-\lambda q+\frac{\lambda q}{h-\widetilde y_i}(x-\widetilde y_i),
      & \widetilde y_i\leq x.
    \end{cases}
  \end{align}
  For sufficiently small $q$, its three successive slopes are positive and weakly decreasing because
  \begin{align*}
    \frac{t_i}{a-\ell}\geq s_i
    \geq\frac{\lambda q}{h-\widetilde y_i},
  \end{align*}
  where the second inequality is equivalent to $z\leq\lambda a+(1-\lambda)h$.
  Thus every $\overline u_i^*$ is continuous, strictly increasing, and concave.

  For every $i$, the construction gives $\overline u_i^*(x_i)=1-q=E\overline u_i^*(p^e)$,
  and hence $c(p^e,\overline u_i^*)=x_i$.\footnote{For $x_i\geq a$, we have $\widetilde y_i-x_i=z-x_i+\lambda(x_i-a)>0$, and hence $x_i\in[a,\widetilde y_i]$. Therefore, the second branch of the definition applies and $\overline u_i^*(x_i)=t_i+s_i(x_i-a)=1-q$ by the definition of $t_i$.}
  It also gives $\overline u_i^*(z)=\lambda \overline u_i^*(a)+(1-\lambda)\overline u_i^*(h)$,
  so that $c\big(\lambda\delta(a\bone)+(1-\lambda)r^e,\overline u_i^*\big)=z$.
  Finally,
  \begin{align*}
    \overline u_i^*(\widetilde y_i)=1-\lambda q
    =\lambda E\overline u_i^*(p^e)+(1-\lambda)\overline u_i^*(h),
  \end{align*}
  which implies $c\big(\lambda p^e+(1-\lambda)r^e,\overline u_i^*\big)=\widetilde y_i$.

\begin{figure}[H]
    \centering
    \setlength{\unitlength}{0.8pt}
    \begin{picture}(440,190)
      \put(55,25){\vector(1,0){190}}
      \put(55,25){\vector(0,1){135}}
      \put(247,20){$\pi_\ell$}
      \put(48,163){$\pi_a$}
      \put(155,168){\makebox(0,0){(a) $x_i<a$}}

      \linethickness{0.4pt}
      \put(55,145){\line(4,-3){160}}
      \qbezier(55,92)(76,58)(96,25)
      \qbezier(55,143)(82,84)(110,25)

      \put(53,23){\circle*{4}}
      \put(108,23){\circle*{4}}
      \put(188,23){\circle*{4}}
      \put(53,90){\circle*{4}}
      \put(53,143){\circle*{4}}

      \put(35,13){$r^e$}
      \put(101,7){$\widehat p$}
      \put(182,7){$p^e$}
      \put(31,87){$p^a$}
      \put(5,139){$\delta(a\bone)$}

      \put(275,105){$c(p^a,u_i^*)=z$}
      \put(275,78){$c(\widehat p,u_i^*)=\widetilde y_i<z$}
    \end{picture}

    \vspace{1.5em}

    \begin{picture}(440,190)
      \put(55,25){\vector(1,0){190}}
      \put(55,25){\vector(0,1){135}}
      \put(247,20){$\pi_\ell$}
      \put(48,163){$\pi_a$}
      \put(155,168){\makebox(0,0){(b) $x_i>a$}}

      \linethickness{0.4pt}
      \put(55,145){\line(4,-3){160}}
      \qbezier(55,92)(83,58)(115,25)
      \qbezier(55,72)(80,48)(105,25)

      \put(53,23){\circle*{4}}
      \put(108,23){\circle*{4}}
      \put(188,23){\circle*{4}}
      \put(53,90){\circle*{4}}
      \put(53,143){\circle*{4}}

      \put(35,13){$r^e$}
      \put(101,7){$\widehat p$}
      \put(182,7){$p^e$}
      \put(31,87){$p^a$}
      \put(5,139){$\delta(a\bone)$}

      \put(275,105){$c(p^a,u_i^*)=z$}
      \put(275,78){$c(\widehat p,u_i^*)=\widetilde y_i>z$}
    \end{picture}
    \caption{The auxiliary lotteries and individual indifference curves. Each point $(\pi_\ell,\pi_a)$ represents the lottery $\pi_\ell\delta(\ell\bone)+\pi_a\delta(a\bone)+(1-\pi_\ell-\pi_a)\delta(h\bone)$, where $\pi_\ell,\pi_a\geq 0$ and $\pi_\ell+\pi_a\leq 1$. The downward-sloping lines are indifference curves induced by $u_i^*$.
    Note that $\widetilde y_i=z$ when $x_i=a$. $\square$} 
    \label{fig:auxiliary-lotteries}
  \end{figure}
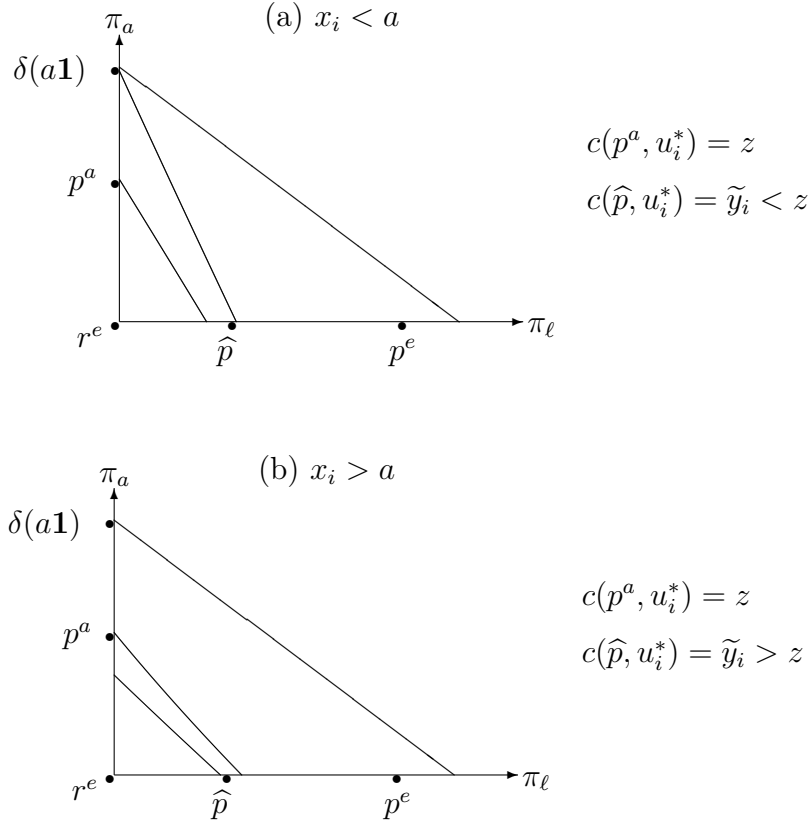

  Now normalize each auxiliary utility by defining
  \begin{align*}
    u_i^*(x)
    =\frac{\overline u_i^*(x)-\overline u_i^*(\underline{x})}
    {\overline u_i^*(\overline{x})-\overline u_i^*(\underline{x})}.
  \end{align*}
  This normalization preserves all the certainty equivalents displayed above.
  Hence $\bu^*=(u_i^*)_{i\in N}\in\mathcal D$ under the full-domain assumption, and for all $i \in N$,
  \begin{align*}
    c(p^e,u_i^*)&=x_i,\\
    c\big(\lambda\delta(a\bone)+(1-\lambda)r^e,u_i^*\big)&=z,\\
    c\big(\lambda p^e+(1-\lambda)r^e,u_i^*\big)&=\widetilde y_i.
  \end{align*}
  Figure~\ref{fig:auxiliary-lotteries} illustrates the construction in the simplex of lotteries supported on $\{\ell\bone,a\bone,h\bone\}$.
  We have $\widetilde y_i-z=\lambda(x_i-a)$ for $i$ with $x_i \geq a$, whereas $\widetilde y_i<z$ for $i$ with $x_i<a$.
  Therefore,
  \begin{align}
    \sum_{i\in N}(\widetilde y_i-z)
    \leq
    \lambda\sum_{i:x_i>a}(x_i-a)-(z-\widetilde y_k)<0.
    \label{14}
  \end{align}

  We claim that
  \begin{align}
    e(\vbx,\succsim^{\bu})=\min_{i\in N}x_i
    \quad\text{for every }\vbx\in \mathcal X\text{ and every }\bu\in\mathcal D.
    \label{15}
  \end{align}
  First consider $\vbx\in\operatorname{int}(X)^N$ and $\bu\in\mathcal D$.
  Continuity and Ex-post Pareto give $e(\vbx,\succsim^{\bu})\geq \min_i x_i$ by the usual argument.
  Suppose to the contrary that this inequality is strict.
  Choose $a\in\operatorname{int}(X)$ so that
  \begin{align*}
    \min_{i\in N}x_i<a<e(\vbx,\succsim^{\bu}).
  \end{align*}
  Using this $\vbx$ and $a$, construct the utility profile $\bu^*\in\mathcal D$ and the lotteries $p^e,r^e\in\Delta^s(\mathcal X)\cap\Delta^e(\mathcal X)$ as above, together with $\lambda\in(0,1)$ and $z\in\operatorname{int}(X)$, so that
  \begin{align*}
    \big(c(p^e,u_i^*)\big)_{i\in N}&=\vbx,\\
    \big(c(\lambda\delta(a\bone)+(1-\lambda)r^e,u_i^*)\big)_{i\in N}&=z\bone.
  \end{align*}
  Let $\vby=(\widetilde y_i)_{i\in N}$ and $\overline{y}_N=n^{-1}\sum_{i\in N}\widetilde y_i$.
  By \eqref{14}, we have $\overline{y}_N<z$.

  We now consider the social ordering $\succsim^{\bu^*}$ associated with the constructed utility profile $\bu^*$.
  By IDL and \eqref{12},
  \begin{align*}
    e(\vbx,\succsim^{\bu^*})
    =e(\vbx,\succsim^{\bu})>a.
  \end{align*}
  Since $c(p^e,u_i^*)=x_i$ for every $i\in N$, all individuals are indifferent between $p^e$ and $\delta(\vbx)$.
  Continuity and Pareto for Equal or No Risk imply $p^e\sim^{\bu^*}\delta(\vbx)$.
  By the definition of the EDE,
  \begin{align*}
    \delta(\vbx)\sim^{\bu^*}\delta\big(e(\vbx,\succsim^{\bu^*})\bone\big)
    \succ^{\bu^*}\delta(a\bone).
  \end{align*}
  Hence $p^e\succ^{\bu^*}\delta(a\bone)$.
  Define
  \begin{align*}
    \widehat{p}&=\lambda p^e+(1-\lambda)r^e,\\
    p^a&=\lambda\delta(a\bone)+(1-\lambda)r^e.
  \end{align*}
  Applying NCI to $p^e\succ^{\bu^*}\delta(a\bone)$ yields
  \begin{align}
    \widehat{p}\succsim^{\bu^*}p^a.
    \label{16}
  \end{align}
  Under Pareto for Equal or No Risk, the comparison in \eqref{16} is equivalent to comparing the EDEs of the vectors of individual certainty equivalents of $\widehat p$ and $p^a$ at the profile $\bu^*$.
  Since
  \begin{align*}
    \big(c(\widehat p,u_i^*)\big)_{i\in N}=\vby
    \quad\text{and}\quad
    \big(c(p^a,u_i^*)\big)_{i\in N}=z\bone,
  \end{align*}
  equation \eqref{16} implies
  \begin{align}
    e(\vby,\succsim^{\bu^*})
    \geq e(z\bone,\succsim^{\bu^*})=z.
    \label{17}
  \end{align}
  On the other hand, a finite sequence of transfers transforms $\vby$ into the equal vector $\overline{y}_N\bone$ while preserving the sum of payoffs.
  By Ex-post Transfer and IDL, applied at the profile $\bu^*$, it follows that
  \begin{align*}
    e(\vby,\succsim^{\bu^*})
    \leq e(\overline{y}_N\bone,\succsim^{\bu^*})
    =\overline{y}_N<z,
  \end{align*}
  contradicting \eqref{17}.
  Thus $e(\vbx,\succsim^{\bu})=\min_i x_i$.
  Since $\bu$ was arbitrary, \eqref{15} holds on $\operatorname{int}(X)^N$ for every profile.
  Continuity extends it to all of $\mathcal X$ for every profile.

  Finally, \eqref{4}, established in Section 6, applies to all $\bu\in\mathcal D$ and all $p,q\in\Delta(\mathcal X)$.
  By \eqref{15}, $e(p,\succsim^{\bu})=e(p,\succsim_M)$ for every $p\in\Delta(\mathcal X)$ and every $\bu\in\mathcal D$.
  Hence \eqref{4} becomes
  \begin{align*}
    p\succsim^{\bu}q
    \Longleftrightarrow
    \min_{i\in N}c\big(e(p,\succsim_M),u_i\big)
    \geq
    \min_{i\in N}c\big(e(q,\succsim_M),u_i\big).
  \end{align*}
  This implies $\succsim=\succsim_M$, as desired. \qed\\

  \section{Independence of the Axioms}

  \subsection*{Proposition 1}

  We first show that Continuity, Ex-ante Pareto, and NCI are mutually independent.
  The examples may be defined on any domain containing the profiles used below; in particular, no full-domain assumption is needed.

  \medskip
  \noindent
  \textbf{Continuity.}
  Define $\succsim_{1L}^{\bu}$ by the lexicographic ordering of individual expected utilities according to the individuals' indices:
  \begin{align*}
    p\succsim_{1L}^{\bu}q
    \quad\Longleftrightarrow\quad
    \big(Eu_1(p),\ldots,Eu_n(p)\big)
    \geq_{\rm lex}
    \big(Eu_1(q),\ldots,Eu_n(q)\big).
  \end{align*}
  Here, $\boldsymbol{a}\geq_{\rm lex}\boldsymbol{b}$ if either $\boldsymbol{a}=\boldsymbol{b}$ or, for the smallest index $j$ such that $a_j\neq b_j$, we have $a_j>b_j$.
  This criterion satisfies Ex-ante Pareto and NCI, but obviously violates Continuity.

  \medskip
  \noindent
  \textbf{Ex-ante Pareto.}
  Define $\succsim_{1P}^{\bu}$ as universal indifference:
  \begin{align*}
    p\sim_{1P}^{\bu}q
    \quad\text{for all }p,q\in\Delta(\mathcal X).
  \end{align*}
  This criterion is continuous and satisfies NCI, but clearly violates Ex-ante Pareto.

  \medskip
  \noindent
  \textbf{Negative Certainty Independence.}
  Define $\succsim_{1N}^{\bu}$ by the representation
  \begin{align*}
    V_{1N}^{\bu}(p)=\sum_{i\in N}\big(Eu_i(p)\big)^2.
  \end{align*}
  Since every normalized expected utility belongs to $[0,1]$, this criterion is continuous and satisfies Ex-ante Pareto.
  It violates NCI.
  For $n=2$ and $X=[0,1]$, at the identical linear-utility profile, let $p=\delta(1,0)$, $\delta(\vbx)=\delta(0.8,0.5)$, and $r=\delta(0,1)$.
  We have
  \begin{align*}
    V_{1N}^{\bu}(p)=1>0.89=V_{1N}^{\bu}(\delta(\vbx)),
  \end{align*}
  whereas
  \begin{align*}
    V_{1N}^{\bu}(0.5p+0.5r)=0.5
    <0.7225
    =V_{1N}^{\bu}\big(0.5\delta(\vbx)+0.5r\big).
  \end{align*}
  Thus NCI is violated.

  \subsection*{Proposition 2}

  We next show that Continuity, Ex-ante Pareto, NCI, and IDL are mutually independent under the full-domain assumption $\mathcal D=\mathcal U^N$.

  \medskip
  \noindent
  \textbf{Continuity.}
  Consider $\succsim_{1L}$ defined above.
  It satisfies Ex-ante Pareto and NCI, but violates Continuity.
  It satisfies IDL.
  Indeed, for any two distinct allocations $\vbx$ and $\vby$, let $j$ be the smallest index for which $x_j\neq y_j$.
  Since $u_j$ is strictly increasing, $u_j(x_j)\geq u_j(y_j)$ if and only if $x_j\geq y_j$.
  Hence, on degenerate lotteries, the lexicographic ordering of
  $\big(u_1(x_1),\ldots,u_n(x_n)\big)$
  according to the individuals' indices is equivalent to the corresponding lexicographic ordering of $(x_1,\ldots,x_n)$ and is therefore independent of the utility profile.

  \medskip
  \noindent
  \textbf{Ex-ante Pareto.}
  Consider the universal-indifference criterion $\succsim_{1P}$ defined above.
  It is continuous and satisfies NCI and IDL, but violates Ex-ante Pareto.

  \medskip
  \noindent
  \textbf{Negative Certainty Independence.}
  Define $\succsim_{2N}^{\bu}$ by the representation
  \begin{align*}
    V_{2N}^{\bu}(p)=\sum_{i\in N}c(p,u_i).
  \end{align*}
  This criterion is continuous and satisfies Ex-ante Pareto.
  It also satisfies IDL because
  \begin{align*}
    V_{2N}^{\bu}(\delta(\vbx))=\sum_{i\in N}x_i
  \end{align*}
  for every profile.
  It violates NCI.
  Let $n=2$, $X=[0,1]$, $u_1(x)=\sqrt{x}$, and $u_2(x)=2x-x^2$, and let
  \begin{align*}
    p&=0.5\delta(0\bone)+0.5\delta(0.3\bone),\\
    r&=\delta(\bone).
  \end{align*}
  Direct calculation gives
  \begin{align*}
    V_{2N}^{\bu}(p)&\simeq0.211866>0.2
    =V_{2N}^{\bu}(\delta(0.1\bone)),\\
    V_{2N}^{\bu}(0.5p+0.5r)&\simeq0.795353
    <0.796718
    \simeq V_{2N}^{\bu}\big(0.5\delta(0.1\bone)+0.5r\big).
  \end{align*}

  \medskip
  \noindent
  \textbf{Invariance on Degenerate Lotteries.}
  Fix $\boldsymbol\beta\in\Delta(N)$ such that at least two coefficients are positive, and define
  \begin{align*}
    p\succsim_{2I}^{\bu}q
    \quad\Longleftrightarrow\quad
    \sum_{i\in N}\beta_i Eu_i(p)
    \geq
    \sum_{i\in N}\beta_i Eu_i(q).
  \end{align*}
  This criterion is continuous and satisfies Ex-ante Pareto and Independence, and hence NCI.
  It violates IDL.
  Let $n=2$, $X=[0,1]$, $\boldsymbol\beta=(1/2,1/2)$, $\vbx=(0.4,0)$, and $\vby=(0,0.5)$.
  At the profile $(u_1(x),u_2(x))=(\sqrt{x},x)$, $\delta(\vbx)\succ_{2I}^{\bu}\delta(\vby)$, whereas at the profile $(u_1'(x),u_2'(x))=(x,\sqrt{x})$, $\delta(\vby)\succ_{2I}^{\bu'}\delta(\vbx)$.

  \subsection*{Proposition 3}

  We next show that Continuity, Ex-post Pareto, Pareto for Equal Risk, and NCI are mutually independent.
  As with Proposition 1, no full-domain assumption is needed.

  \medskip
  \noindent
  \textbf{Continuity.}
  The lexicographic criterion $\succsim_{1L}$ defined above satisfies Ex-ante Pareto and NCI, and therefore satisfies both Ex-post Pareto and Pareto for Equal Risk.
  As shown above, it violates Continuity.

  \medskip
  \noindent
  \textbf{Ex-post Pareto.}
  For each profile $\bu$, let
  \begin{align*}
    W_{3E}^{\bu}(\vbx)
    =\frac{1}{n}\sum_{i\in N}u_i(x_i)
    -K\sum_{i\in N}(x_i-\overline{x}_N)^2,
    \qquad
    \overline{x}_N=\frac{1}{n}\sum_{i\in N}x_i,
  \end{align*}
  where $K>0$, and define $\succsim_{3E}^{\bu}$ by expected utility with the vNM index $W_{3E}^{\bu}$.
  This criterion is continuous and satisfies Independence, and hence NCI.
  On equal lotteries, the inequality-penalty term vanishes and the criterion is represented by the average of the individual expected utilities, so it satisfies Pareto for Equal Risk.
  It may violate Ex-post Pareto.
  Let $n=2$, $X=[0,1]$, $K=2$, and $u_1(x)=u_2(x)=x$.
  For $\vbx=(1,0.1)$ and $\vby=(0,0)$, we have $\vbx\gg\vby$, but
  \begin{align*}
    W_{3E}^{\bu}(\vbx)
    =0.55-2(0.45^2+0.45^2)
    =-0.26
    <0=W_{3E}^{\bu}(\vby).
  \end{align*}

  \medskip
  \noindent
  \textbf{Pareto for Equal Risk.}
  Define $\succsim_{3P}$ by expected total monetary payoff:
  \begin{align*}
    p\succsim_{3P}q
    \quad\Longleftrightarrow\quad
    \int_{\mathcal X}\sum_{i\in N}x_i\,dp(\vbx)
    \geq
    \int_{\mathcal X}\sum_{i\in N}x_i\,dq(\vbx).
  \end{align*}
  This criterion is continuous, satisfies Independence and hence NCI, and satisfies Ex-post Pareto.
  It violates Pareto for Equal Risk.
  Let $X=[0,1]$ and $u_i(x)=\sqrt{x}$ for every $i$, and consider
  \begin{align*}
    p^e=\delta(0.4\bone),
    \qquad
    q^e=0.5\delta(0\bone)+0.5\delta(\bone).
  \end{align*}
  Every individual strictly prefers $p^e$ to $q^e$ because $\sqrt{0.4}>0.5$, whereas $\succsim_{3P}$ ranks $q^e$ above $p^e$ because $0.5>0.4$.

  \medskip
  \noindent
  \textbf{Negative Certainty Independence.}
  The criterion $\succsim_{1N}$ represented by $\sum_i(Eu_i(p))^2$ is continuous and satisfies Ex-ante Pareto, and hence satisfies Ex-post Pareto and Pareto for Equal Risk.
  As shown above, it violates NCI.

  \subsection*{Proposition 4}

  We next show that the five axioms in Proposition 4 are mutually independent.
  Throughout this subsection, $\mathcal D=\mathcal U^N$, as assumed in Proposition 4.

  \medskip
  \noindent
  \textbf{Continuity.}
  For every $p\in\Delta(\mathcal X)$ and $i\in N$, let
  \begin{align*}
    z_i^{\bu}(p)=c\big(e(p,\succsim_M),u_i\big),
  \end{align*}
  and let $z_{(1)}^{\bu}(p)\leq\cdots\leq z_{(n)}^{\bu}(p)$ denote the ordered components of this vector.
  Define $\succsim_L^{\bu}$ by the leximin ordering
  \begin{align*}
    p\succsim_L^{\bu}q
    \quad\Longleftrightarrow\quad
    \big(z_{(1)}^{\bu}(p),\ldots,z_{(n)}^{\bu}(p)\big)
    \geq_{\rm lex}
    \big(z_{(1)}^{\bu}(q),\ldots,z_{(n)}^{\bu}(q)\big).
  \end{align*}
  Thus, $\succsim_L$ is the leximin extension of the maximin CEDE criterion: it preserves every strict comparison made by $\succsim_M$ and lexicographically resolves its indifferences.
  It is immediate that this criterion satisfies Pareto for Equal or No Risk, NCI, IDL, and Ex-post Transfer, and violates Continuity.

  \medskip
  \noindent
  \textbf{Pareto for Equal or No Risk.}
  The generalized-Gini CEDE criterion $\succsim_G$ in \hyperref[ex:generalized-gini]{Example 1} satisfies Continuity, NCI, and IDL.
  It also satisfies Ex-post Transfer because its EDE function $g_{\boldsymbol\gamma}$ weakly increases under every rank-preserving Pigou--Dalton transfer.
  As shown in Section 6, however, it violates Pareto for Equal or No Risk.

  \medskip
  \noindent
  \textbf{Negative Certainty Independence.}
  Fix $\boldsymbol{\gamma}\in\Delta(N)$ with $\gamma_1>\cdots>\gamma_n>0$ and define
  \begin{align*}
    V_R^{\bu}(p)
    =g_{\boldsymbol{\gamma}}\Big(\big(c(g_{\boldsymbol{\gamma}}(p),u_i)\big)_{i\in N}\Big),
  \end{align*}
  where $g_{\boldsymbol\gamma}$ is the generalized-Gini EDE function defined in \hyperref[ex:generalized-gini]{Example 1}.
  Let $\succsim_R^{\bu}$ be represented by $V_R^{\bu}$.
  This criterion is continuous and satisfies IDL because
  \begin{align*}
    V_R^{\bu}(\delta(\vbx))=g_{\boldsymbol\gamma}(\vbx),
  \end{align*}
  independently of $\bu$.
  The same equality and the transfer property of $g_{\boldsymbol\gamma}$ imply Ex-post Transfer.
  Pareto for Equal or No Risk follows from the strict monotonicity of $g_{\boldsymbol\gamma}$: for equal prospects, the relevant vector is the vector of individual certainty equivalents; for degenerate lotteries, it is the allocation itself.

  The criterion need not satisfy NCI.
  Let $n=2$, $X=[0,1]$, $\boldsymbol\gamma=(0.55,0.45)$, $u_1(x)=\sqrt{x}$, and $u_2(x)=2x-x^2$.
  Consider the equal prospects
  \begin{align*}
    p&=0.5\delta(0\bone)+0.5\delta(0.3\bone),\\
    r&=0.5\delta(0.6\bone)+0.5\delta(0.7\bone).
  \end{align*}
  Direct calculation gives
  \begin{align*}
    V_R^{\bu}(p)&\simeq0.102840>0.1
    =V_R^{\bu}(\delta(0.1\bone)),\\
    V_R^{\bu}\big(0.5p+0.5r\big)&\simeq0.313433
    <0.315369
    \simeq V_R^{\bu}\big(0.5\delta(0.1\bone)+0.5r\big).
  \end{align*}
  Thus the implication required by NCI is reversed after mixing.

  \medskip
  \noindent
  \textbf{Invariance on Degenerate Lotteries.}
  Fix a coefficient vector $\boldsymbol\beta\in\Delta(N)$ such that $\beta_i>0$ for every $i\in N$.
  Define $\succsim_I$ as follows.
  At every identical-preference profile, let $\succsim_I^{\bu}=\succsim_M^{\bu}$.
  At every other profile, let
  \begin{align*}
    p\succsim_I^{\bu}q
    \quad\Longleftrightarrow\quad
    \sum_{i\in N}\beta_i Eu_i(p)
    \geq
    \sum_{i\in N}\beta_i Eu_i(q).
  \end{align*}
  At each profile the criterion is continuous and satisfies NCI.
  It also satisfies Pareto for Equal or No Risk: this follows from the properties of $\succsim_M$ at identical-preference profiles and from the positivity of every $\beta_i$ at all other profiles.
  Ex-post Transfer holds because the axiom applies only at identical-preference profiles, where the criterion coincides with $\succsim_M$.

  The criterion violates IDL.
  Let $n=2$, $X=[0,1]$, $\boldsymbol\beta=(1/2,1/2)$, $\vbx=(1,0)$, and $\vby=(0.2,0.2)$.
  At the identical linear-utility profile $u_1(x)=u_2(x)=x$, maximin gives
  \begin{align*}
    \delta(\vby)\succ_I^{\bu}\delta(\vbx).
  \end{align*}
  At the nonidentical profile $u_1(x)=x$ and $u_2(x)=\sqrt{x}$, however,
  \begin{align*}
    \frac{u_1(1)+u_2(0)}{2}
    =0.5
    >\frac{u_1(0.2)+u_2(0.2)}{2},
  \end{align*}
  and hence $\delta(\vbx)\succ_I^{\bu}\delta(\vby)$.

  \medskip
  \noindent
  \textbf{Ex-post Transfer.}
  Fix $d\in N$ and define the dictatorial criterion
  \begin{align*}
    p\succsim_D^{\bu}q
    \quad\Longleftrightarrow\quad
    Eu_d(p)\geq Eu_d(q).
  \end{align*}
  This criterion obviously satisfies Continuity, Pareto for Equal or No Risk, NCI, and IDL, but violates Ex-post Transfer. \qed\\

\end{appendices}

\end{document}